\documentclass[aip,pop,preprint,superscriptaddress,amsmath,amsfonts,amssymb]{revtex4-1}
\usepackage{graphicx}
\usepackage{amsfonts}
\usepackage{amsmath}
\usepackage{bm}
\usepackage[utf8]{inputenc}
\usepackage[T1]{fontenc}
\usepackage[colorlinks,linkcolor=blue,citecolor=blue,bookmarks,pdfstartview=FitH]{hyperref}
\usepackage{graphics}
\newcommand{\idest}{{\em i.e.}, }

\newcommand{\figref}[1]{FIG.\ref{#1}}

\begin{document}

\title{Shear Alfvén and acoustic continuum in general axisymmetric toroidal geometry}

\author{Matteo Valerio Falessi}
\email{matteo.falessi@enea.it}
\affiliation{ENEA, Fusion and Nuclear Safety Department, C. R. Frascati,\\ Via E. Fermi 45, 00044 Frascati (Roma) (Italy)}
\affiliation{National Institute for Nuclear Physics, Rome Department,\\ P.le A. Moro 5, 00185 Roma (Italy)}
\author{Nakia Carlevaro}
\affiliation{ENEA, Fusion and Nuclear Safety Department, C. R. Frascati,\\ Via E. Fermi 45, 00044 Frascati (Roma) (Italy)}
\affiliation{Consorzio RFX, Corso Stati Uniti 4, 35127 Padova (Italy)}
\author{Valeria Fusco}
\affiliation{ENEA, Fusion and Nuclear Safety Department, C. R. Frascati,\\ Via E. Fermi 45, 00044 Frascati (Roma) (Italy)}
\author{Gregorio Vlad}
\affiliation{ENEA, Fusion and Nuclear Safety Department, C. R. Frascati,\\ Via E. Fermi 45, 00044 Frascati (Roma) (Italy)}
\author{Fulvio Zonca}
\affiliation{ENEA, Fusion and Nuclear Safety Department, C. R. Frascati,\\ Via E. Fermi 45, 00044 Frascati (Roma) (Italy)}
\affiliation{Institute for Fusion Theory and Simulation and Department of Physics,
Zhejiang University, Hangzhou 310027 (China)}

\begin{abstract}
The equations describing the continuous spectrum of shear Alfvén and ion sound waves propagating along magnetic field lines are introduced and solved in the ballooning space for general geometry in the ideal MHD limit. This approach is equivalent to earlier analyses by Chu \emph{et al.} 1992 [Phys. Fluids B \textbf{4}, 3713 (1992)] but the present formulation in the ballooning space allows to readily extend it to include gyrokinetic and three-dimensional equilibrium effects. In particular, following Chen and Zonca 2017 [Phys. Plasmas \textbf{24}, 072511 (2017)], the MHD limit is adopted to illustrate the general methodology in a simple case, and the equations are solved within the framework of Floquet and Hill's equation theory. The connection of shear Alfvén and ion sound wave continuum structures to the generalized plasma inertia in the general fishbone like dispersion relation is also illustrated and discussed. As an application, the continuous frequency spectrum is calculated for a reference equilibrium of the Divertor Tokamak Test facility. The results are compared with those obtained by the MARS code adopting the standard methodology, demonstrating excellent agreement.
\end{abstract}
\maketitle

\section{Introduction}
\label{sec:orge5b856d}
Computing the shear Alfv\'en wave (SAW) continuous frequency spectrum\cite{barston64,sedlacek71,grad69,uberoi72,tataronis73,hasegawa74,chen74a,chen74b,dewar74,appert74,goedbloed75,pao75,chance77} 
in general toroidal geometry is important because it 
determines mode structures and dispersive properties of Alfv\'enic fluctuations excited by energetic particles (EPs) in fusion devices,
which, in turn, play crucial role in determining the plasma collective behaviors and eventually may affect fusion performance\cite{chen16}.
Due to equilibrium magnetic field curvature\cite{pogutse78,dippolito80,kieras82,cheng85,cheng86}, SAW and ion sound wave (ISW) continuous frequency 
spectrum are coupled and, thus, their
structures should be considered self-consistently, taking into account realistic equilibrium magnetic field geometry and plasma nonuniformity.

In numerical applications adopting realistic equilibrium reconstruction, the typical approach to the coupled SAW and ISW spectra is to Fourier decompose
the fluctuation structure in general poloidal and toroidal angles and to compute the 
null space (kernel) of the matrix of the highest order radial derivative\cite{cheng86}, e.g., of the radial plasma displacement. This is completely equivalent to solve for the radial singular structures
of the coupled SAW and ISW propagating along the field lines, described by two second order coupled differential equations\cite{chu92}. 
In particular, the latter method has been adopted to compute the SAW and ISW spectra in DIII-D\cite{chu92}, 
focusing on both shaping and finite pressure effects on the continuous spectrum structures at low frequency\cite{chu92,huysmans95,goedbloed98,vanderholst00}.
If the ideal MHD plasma description is adopted, again, the two approaches described above are mathematically equivalent, although it has been pointed out in Ref. \citenum{chu92} that the second method has advantages when computing the high toroidal mode number continuous spectra.

Kinetic description becomes necessary at short wavelengths and/or low frequencies to account for finite parallel electric field and wave damping.
For the SAW-ISW coupling at low frequency and corresponding frequency gaps\cite{chu92,huysmans95,goedbloed98,vanderholst00} with possible discrete modes localized therein\cite{gorelenkov07a,gorelenkov07b,gorelenkov09}, the need of kinetic theory for the damping assessment was pointed out in Refs. \citenum{chavdarovski09,zonca10,chavdarovski14} and motivated gyrokinetic stability studies of low-frequency Alfv\'enic fluctuations exctited by EPs with realistic geometry and plasma profiles\cite{zhang16,liu17,Bierwage2017}. The radial singular structures characterizing the SAW and ISW continuous spectra are most easily isolated\cite{pegoraro86,newcomb90} adopting the 
ballooning mode representation\cite{connor78}, or the mode structure decomposition approach\cite{lu12}, which applies for arbitrary mode numbers and reduces to the ballooning representation in the high mode number limit. Within this framework, the general description of SAW and ISW using gyrokinetic theory has been
given in Refs. \citenum{chen16,zonca14a,zonca14b}, and allows to include thermal plasma\cite{zonca96b,zonca98,zonca99,Lauber2013,bowden15,Lauber2018} as well as EPs induced\cite{zonca96a,zonca99,Lauber2013,zonca14b,bowden15}  kinetic effects in the analysis of the continuous spectra.

This premise provides a motivation of the present work, which is to show that the coupled second order differential equations describing SAW and ISW propagation along equilibrium magnetic field lines for general geometry are most easily expressed and solved in the ballooning space. In the ideal MHD limit, as already stated, this is equivalent to the approach adopted in Ref. \citenum{chu92}. The present methodology, however, is readily extended to include gyrokinetic description\cite{chen16,zonca14a,zonca14b} and 3D geometry\cite{dewar83}. 
Furthermore, this formulation not only allows us to discuss the structures of SAW and ISW continuous spectra in general toroidal geometry, but it provides the simplest and most direct way of computing the {\em generalized inertia} as an ingredient of the general fishbone-like dispersion relation (GFLDR)\cite{chen16,zonca14a,zonca14b}; that is, the unified framework for describing Alfvénic fluctuations excited by EPs in tokamaks. In Sec. \ref{sec:org90d93a1}, in order to establish a link to prior work more closely, we take the ideal MHD limit\cite{chen17} of the general description\cite{chen16,zonca14a,zonca14b} and we write the coupled SAW and ISW equations for radial singular structures in general tokamak geometry. Along with providing a synthetic summary of prior theoretical analyses, which are applied here to realistic plasma equilibria, we illustrate the link of  the generalized inertia in the GFLDR to the structures of the continuous spectra and show how it is computed\cite{chen16,zonca14a,zonca14b,chen17}. Furthermore, the important role of polarization of physical fluctuations is analyzed with emphasis on its relevance for assessing their effective absorption by resonantly excited radial singular structures of the continuous spectrum. We then analyze the general form of the obtained equations, collocating that within the framework of Floquet theory\cite{floquet1883equations} and Hill's equation\cite{hill1886part,magnus2013hill,Denk1995}. As original application, in Sec. \ref{sec:org78c95f5} we solve for the SAW and ISW continuous spectra for a reference equilibrium of the Divertor Tokamak Test (DTT) facility\cite{albanese17,albanese19}. We discuss both (artificially) decoupled and (realistic) coupled cases to illustrate features of the obtained SAW and ISW spectra and directly compare results of the present approach with those obtained by the MARS\cite{Bondeson1992} code, which is equivalent to computing the null space of the matrix of the highest order radial derivative\cite{cheng86} of the plasma radial displacement. In particular, we are able to identify the role of various equilibrium geometry effects in SAW-SAW, ISW-ISW and SAW-ISW couplings and corresponding structures in the continuous spectra. All these detailed features are naturally accounted for in the present formulation and can be included as boundary conditions for calculating parallel mode structures and GFLDR\cite{zonca14a,zonca14b} dispersion relations in general Tokamak geometry. Doing so, however, is beyond the scope of the present work and will be done elsewhere. Further extensions to gyrokinetic analyses and 3D geometry are discussed in Sec \ref{sec:org6275bbc}, where concluding remarks are provided along with considerations about future perspectives.

\section{Theoretical framework}
\label{sec:org90d93a1}
The general theoretical framework is discussed in Ref. \citenum{chen16}, based on the detailed derivations given by Zonca and Chen in Refs. \citenum{zonca14a,zonca14b}. There, the general equations governing SAW and ISW fluctuation in low-$\beta$ magnetized fusion plasmas are given by quasineutrality condition and vorticity equation, which are derived from the gyrokinetic equation\cite{frieman1982nonlinear}. Meanwhile, Ref. \citenum{chen16} also provides a thorough discussion of how gyrokinetic vorticity equation and quasineutrality condition yield the proper reduced MHD limit when appropriate \cite{brizard1994eulerian,qin1998gyrokinetic,Qin1999}. The reduced MHD limiting form is also derived by the standard approach in Ref. \citenum{chen17}. In the present application to general tokamak equilibria, we merely assume the aforementioned theoretical framework without derivation and only provide the essential elements for introducing the notation and solving the relevant equations. The general features of the mode structure decomposition of fluctuations in toroidal plasmas\cite{lu12} are also discussed in Refs. \citenum{zonca14a,zonca14b} along with its usefulness to generally describe continuous spectra.
\subsection{Notation and fundamental equations}
\label{sec:org6682307}
In this work we assume an axisymmetric equilibrium magnetic field \(\boldsymbol{B}_{0}\) expressed in flux coordinates \((r,\theta,\varphi)\):
\begin{equation}
\label{eq:27}
\boldsymbol { B } _ { 0 } = F ( \psi ) \boldsymbol { \nabla } \varphi + \boldsymbol{\nabla} \varphi \times \boldsymbol{ \nabla } \psi\;,
\end{equation}
where \(r(\psi)\) is a radial-like flux coordinate, $\varphi$ is the physical toroidal angle and the angular coordinate \(\theta\) can be chosen such that the Jacobian \(J = ( \boldsymbol { \nabla } \psi \times \boldsymbol { \nabla } \theta \cdot \boldsymbol { \nabla } \varphi ) ^ { - 1 }\) has a convenient expression. Following the common practice, we also introduce the straight field line toroidal angle $\zeta$, defined such that the safety factor
\begin{equation}
q = \frac{\bm B_0 \cdot \bm \nabla \zeta}{\bm B_0 \cdot \bm \nabla \theta} = q(r)
\end{equation}
is a flux function. We introduce the leading order perpendicular plasma displacement as usual,
\begin{equation}
\delta \bm \xi_{\perp} = \frac{c}{B_0} \bm b \times \bm \nabla \Phi_s \; , \label{eq:xiperp0}
\end{equation}
with $\Phi_s$ being the perturbed stream function\cite{chen17}. The fluctuating field 
\begin{align}\nonumber
\Phi_s (r,\theta,\zeta) = \sum_{m} \exp ( i n \zeta - i m \theta ) \Phi_{s m}(r)
\end{align}
expressed as a Fourier series, where $m$ and $n$ represent the poloidal and toroidal mode numbers, respectively, can be decomposed as\cite{lu12}:
\begin{eqnarray}
\label{eq:deco}  
\Phi_s (r,\theta,\zeta) & = & 2\pi \sum_{\ell \in \mathbb{Z}} e^{in\zeta-inq(\theta-2\pi\ell)}  \hat \Phi_s (r,\theta-2\pi\ell)= 
\nonumber \\ & = &  \sum_{m \in \mathbb{Z}} e^{in\zeta-i m\theta} \int d\vartheta e^{i(m-nq)\vartheta} \hat \Phi_s (r,\vartheta)  \; .  \label{eq:msd}
\end{eqnarray}
Here, time dependences are left implicit for simplicity of notation, and $\vartheta$ represents the extended poloidal angle\cite{connor78} coordinate following equilibrium magnetic field lines. Periodic poloidal angle-$\theta$ dependences of equilibrium quantities are replaced by the same periodic dependences on $\vartheta$. However, while fluctuations must be periodic in $\theta$ space for physical reasons, they are generally not periodic in $\vartheta$, as shown in Eq. (\ref{eq:msd}). The radial singular structures corresponding to the continuous spectra are obtained from the limiting forms of, respectively, vorticity and pressure equations for $|\vartheta|\rightarrow \infty$\cite{zonca14a,zonca14b,chen17}. Following Refs. \citenum{chen16,zonca14a,zonca14b}, it is possible to derive the expressions describing the structure of the continuum modes in flux coordinates. In particular, assuming $JB_0^2$ constant on a flux surface, yielding the so called ``Boozer coordinates'', we obtain\cite{chen17}:

\begin{subequations}\label{eq:28}
\begin{align}
\left( \partial _ { \vartheta } ^ { 2 } - \frac { \partial _ { \vartheta } ^ { 2 } | \bm\nabla r | } { | \bm\nabla r | } + \frac { \omega ^ { 2 } J ^ { 2 } B _ { 0 } ^ { 2 } } { v _ { A } ^ { 2 } } \right) y _ { 1 } &= ( 2 \Gamma \overline { \beta } ) ^ { 1 / 2 } \kappa _ { g } \frac { J ^ { 2 } B _ { 0 } \overline { B } _ { 0 } } { q R _ { 0 } } \frac { s \vartheta } { | s \vartheta | } y _ { 2 }\;, \\
   \left( 1 + \frac { c _ { s } ^ { 2 } } { \omega ^ { 2 } } \frac { 1 } { J ^ { 2 } B _ { 0 } ^ { 2 } } \partial _ { \vartheta } ^ { 2 } \right) y _ { 2 }&= ( 2 \Gamma \overline { \beta } ) ^ { 1 / 2 } \kappa _ { g } \frac { \overline { B } _ { 0 } } { B _ { 0 } } q R _ { 0 } \frac { s \vartheta } { | s \vartheta | } y _ { 1 }                                                                                     \;,
\end{align}
\end{subequations}
where
\begin{equation}
\label{eq:17}
y _ { 1 } \equiv \frac { \hat { \phi } _ { s } } { \left( \overline{\beta} q ^ { 2 } \right) ^ { 1 / 2 } } \frac { c k_{\vartheta} } { \bar{B}_{0} R _ { 0 } }\;, \qquad y_{2} \equiv i \frac{\delta \hat{P}_{comp}}{(2 \Gamma)^{1/2}P_{0}}\;,
\end{equation}
$\hat \phi_s (r,\vartheta) \equiv |s \vartheta| |\bm \nabla r| \hat \Phi_s(r,\vartheta)$, $s=rq'/q$ is the magnetic shear, $k_{\vartheta} = - n q/r$, and $\delta \hat P_{comp}(r, \vartheta)$ is the representation of the compressional component of the pressure perturbation according to Eq. (\ref{eq:msd})\cite{chen17}. Furthermore, $\kappa_g$ is the geodesic curvature in Boozer coordinates:
\begin{equation}
\label{eq:5}
\kappa _ { g } = - \frac { 1 } { J B _ { 0 }  } \frac { F } { | \bm\nabla \psi | } \frac { \partial } { \partial \vartheta } \ln{B _ { 0 }}\;,
\end{equation}
$\bar{\beta}= 8\pi \Gamma P_0/\bar B_0^2$, \(\bar{B}_{0}\) is the magnetic field on axis, $r = a \rho_{tor}$, where $\rho_{tor}$ is defined, as usual, as a function of the normalized toroidal magnetic flux and $a$ is the minor radius of the torus. Note, again, that periodic equilibrium field dependences on $\theta$ are replaced by the same periodic dependences on $\vartheta$ in the ballooning space.

These equations describe SAWs and ISW continuous spectra  coupled by equilibrium non-uniformities and geodesic curvature. Both type of waves propagate along equilibrium magnetic field lines and, therefore, radial equilibrium nonuniformities are accounted for by $r$, which enters in Eqs. (\ref{eq:28}) as a parameter. Note also that the continuous spectrum does not depend on magnetic shear although it formally appears in Eqs. (\ref{eq:28}) and (\ref{eq:17}). This can be understood since the continuous spectrum is a local feature of singular radial fluctuation structures. More in depth discussion about this point and what happens for vanishing magnetic shear can be found in Refs. \citenum{chen16,zonca14a,zonca14b}. Equations (\ref{eq:28}), obtained in the MHD limit\cite{chen17}, are valid for general tokamak equilibrium geometry and can be used to calculate the coupled SAW and ISW continuous spectra for cases of practical interest, such as the DTT reference scenario\cite{albanese17,albanese19} introduced in Sec. \ref{sec:referencescenario}. These equations are readily extended to 3D plasma equilibria, such as stellarators, thanks to the generality of ballooning representation\cite{dewar83}. The same formulation can be adopted within a linear gyrokinetic description of coupled SAW and ISW continua, in which case Eqs. (\ref{eq:28}) are replaced by the short radial scale gyrokinetic vorticity equation, \idest Eq. (6) of Ref. \citenum{zonca14b}, and the quasineutrality condition, \idest Eq. (36) of the same work. This theoretical framework is similar to the description of electron wave-packets in a 1D periodic lattice, see e.g. Ref. \citenum{kittel1976introduction}. The interesting analogy\cite{chen16} is due to the fact that the governing equations of low-frequency plasma waves in magnetized plasmas describe the propagation of wave-packets along magnetic field lines with periodicity given by the plasma connection length. Therefore, this correspondence is independent of the axisymmetry of the equilibrium magnetic field and/or the fluid/kinetic nature of the theory. In particular, Eqs. (\ref{eq:17}), are Schrödinger-like and produce structures analogue to electronic bands in solid-state physics. Further discussion of this interesting analogy is given in Refs. \citenum{chen16,zonca14a,zonca14b}.In the following, we will focus on the solution of Eqs. (\ref{eq:28}). By direct inspection, it is readily noted that, adopting Boozer coordinates, ISW continuous spectrum is more simply represented than SAW continuous spectrum, due to the modulation term $\propto \partial_\vartheta^2 |\bm \nabla r|/ |\bm \nabla r|$. Since there is usually more interest in the SAW than in the ISW branch due to the stronger kinetic damping of the latter, we can rewrite Eqs. (\ref{eq:28}), introducing an appropriate angular coordinate to express the SAW frequency continuum more transparently; \idest
\begin{equation}
\label{eq:1}
\eta \equiv 2 \pi \frac{\int _ { 0 } ^ { \vartheta } d \vartheta ^ { \prime } |\bm \nabla r|^{-2}}{\int _ { 0 } ^ { 2 \pi } d \vartheta ^ { \prime } |\bm\nabla r|^{-2}}\;.
\end{equation}
By construction, the Jacobian \(J_{\eta}\) of this new set of coordinates, dubbed continuum flux coordinates (CFC), is such that \(J_{\eta} B_{0}^{2}/|\bm\nabla r|^{2}\) is a flux function. We obtain the following expression for the derivative along \(\eta\) (at constant $\psi$):
\begin{equation}
\label{eq:14}
| \bm\nabla r | ^ { 2 } \mathcal{C} \left. \frac { \partial } { \partial \vartheta } \right| _ { \psi } = \left. \frac { \partial } { \partial \eta } \right| _ { \psi }\;,
\end{equation}
where $\mathcal{C} = (2 \pi)^{-1} \int _ { 0 } ^ { 2 \pi } d \vartheta ^ { \prime } \left| \bm\nabla r \right| ^ { - 2 }$. The Jacobians in the two set of coordinates are related by the following expression:
\begin{equation}
\label{eq:21}
J_{\eta} = J \mathcal{C} | \bm\nabla r | ^ { 2 }\;.
\end{equation}
Introducing the normalizations:
\begin{equation}
\label{eq:22}
g_{1} = \frac{1}{|\bm \nabla r | }y_{1}, \quad g_{2} = \frac{c_{s}}{\omega q R_{0} | \bm\nabla r |} \frac{|s \vartheta|}{s \vartheta}y_{2}\;,
\end{equation}
we can re-write Eqs. (\ref{eq:28}) as
\begin{subequations}\label{eq:29sis}
\begin{align}
 \left( \partial _ { \eta } ^ { 2 } + \frac { \omega ^ { 2 } J _ { \eta } ^ { 2 } B _ { 0 } ^ { 2 } } { v _ { A } ^ { 2 } } \right) g _ { 1 } &=  ( 2 \Gamma \overline { \beta } ) ^ { 1 / 2 } \kappa _ { g } \frac { \omega J _ { \eta } ^ { 2 } B _ { 0 } \overline { B } _ { 0 }  } { c _ { s } } g _ { 2 }\;, \\
  \left( \partial _ { \eta } ^ { 2 } - | \bm\nabla r | \partial _ { \eta } ^ { 2 } | \bm\nabla r | ^ { - 1 } + \frac { \omega ^ { 2 } J _ { \eta } ^ { 2 } B _ { 0 } ^ { 2 } } { c _ { s } ^ { 2 } } \right) g _ { 2 } &=  ( 2 \Gamma \overline { \beta } ) ^ { 1 / 2 } \kappa _ { g } \frac { \omega J _ { \eta } ^ { 2 } B _ { 0 } \overline { B } _ { 0 }  } { c _ { s }  } g _ { 1 }\;,\label{eq:29}
\end{align}
\end{subequations}
respectively, where the geodesic curvature in this new set of coordinates reads:
\begin{equation}
\label{eq:7}
\kappa _ { g } = - \frac { 1 } { J _ { \eta } B_{0}^{2}} \frac { F } { | \bm\nabla \psi | }  \frac { \partial } { \partial \eta } B _ { 0 }\;.
\end{equation}
Equations (\ref{eq:29sis}) are completely equivalent to Eqs. (\ref{eq:28}) and will be  solved in the following, illustrating the properties of coupled SAW and ISW continuous frequency spectra in general tokamak plasma equilibria. They are in the form of two coupled second order differential equations with periodic coefficients. Therefore, all known results of Floquet theory\cite{floquet1883equations} can be applied. These are summarized in the remaining part of Sec. \ref{sec:org90d93a1} for the reader's convenience.
\subsection{Linear differential equations with periodic coefficients}
\label{sec:periodic}
\subsubsection{Floquet theory}
\label{sec:periodicflo}
Equations (\ref{eq:29sis}) are a linear system of second order differential equations with periodic coefficients. Therefore, after re-writing it as four first order differential equations,  we know from Floquet theory\cite{floquet1883equations} that it must have solutions in the form:
\begin{equation}
\label{eq:34}
\mathbf{x}_{i} =e ^ {i \nu_{i} \eta } \mathbf { P }_{i} ( \eta )\;,
\end{equation}
where \(\mathbf{P}_{i}\) is a \(2 \pi\)-periodic function, $i = 1,2,3,4$ and the $\nu_{i}$ are complex numbers called characteristic Floquet exponents of Eqs. (\ref{eq:29sis}). Finding the solutions of Eqs. (\ref{eq:29sis}) for a given value of $r$ and, therefore, calculating \(\nu_{i}\) for each \(\omega\) value is equivalent to obtaining the (local) dispersion curves of the system:
\begin{equation}
\label{eq:33}
\nu_{i} = \nu_{i}(\omega,r)\;.
\end{equation}
We emphasize that this relation involves only local quantities and describes wave packets propagating along magnetic field on a given flux surface. Recalling the physical meaning of the derivative along the $\vartheta$ coordinate\cite{chen16,zonca14a,zonca14b}, we can relate the characteristic Floquet exponents to $\nabla_{\parallel}\equiv (JB_0)^{-1} \partial_\vartheta = (J_\eta B_0)^{-1} \partial_\eta$, and, in particular, to the toroidal and poloidal mode numbers $n,m$:
\begin{equation}
\label{eq:356}
\nu_{i} ^ { 2 } ( \omega , r ) = ( n q ( r ) - m ) ^ { 2 }\;.
\end{equation}
Equation (\ref{eq:356}) readily follows from the definition of $\nabla_\parallel$ and the mode structure decomposition of Eq. (\ref{eq:msd})\cite{chen16,zonca14a,zonca14b,chen17}. Therefore, integrating Eqs. (\ref{eq:29sis}) for different $\omega$ values and using Eqs. (\ref{eq:33}) and (\ref{eq:356}), it is possible to calculate the continuous frequency spectrum for every value of the toroidal mode number. In particular, Eq. (\ref{eq:356}) explicitly shows that continuous spectra can be computed for arbitrary $n$, once the dispersion curves $\nu_i = \nu_i (\omega, r)$ are given. Extensions to 3D/stellarator geometry and gyrokinetic theory would be obtained in the same way from the more general governing equations as briefly discussed in Sec. \ref{sec:org6682307}.

\begin{figure}[!htpb]
\centering
\includegraphics[width=0.8\textwidth]{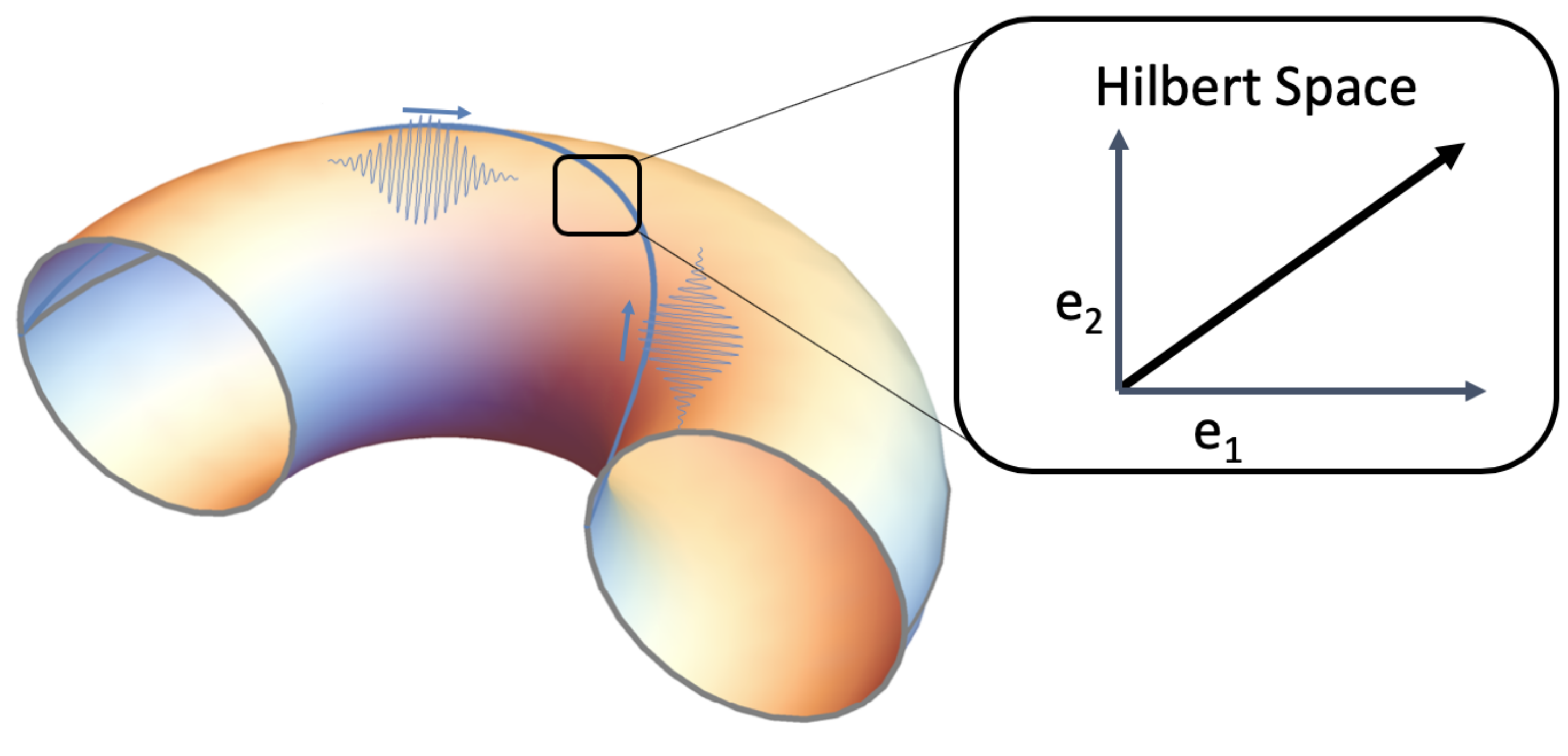}
\caption{Parallel propagation of radial singular structures of the continuous spectrum and their polarization in the orthogonal Hilbert space, which is a ``virtual'' (non-physical) linear space where Floquet solutions $[g_1^{(i)}(\eta;\nu_i,r), g_2^{(i)}(\eta;\nu_i,r)]$ of Eqs. (\ref{eq:29sis}) are represented.\label{fig_zero}}
\end{figure}

For given $\nu_i(\omega,r)$; \idest for given dispersion curves of Eqs. (\ref{eq:29sis}), it is possible to calculate the corresponding vector solutions; \idest eigenvectors of the fundamental matrix $\mathbf{X}(\vartheta)$, that is a matrix-valued function whose columns are linearly independent solutions of Eqs. (\ref{eq:29sis}). Furthermore, since the considered problem is linear, we can generally write the normalized eigenvector solution corresponding to $\nu_i(\omega,r)$ as
\begin{eqnarray}
  g_1^{(i)}(\eta;\nu_i,r) & = & e_1^{(i)}(\nu_i,r) \hat g_1^{(i)}(\eta;\nu_i,r) \; , \nonumber \\
g_2^{(i)}(\eta;\nu_i,r) & = & e_2^{(i)}(\nu_i,r) \hat g_2^{(i)}(\eta;\nu_i,r) \; . \label{eq:eigencont}
\end{eqnarray}
Here, without loss of generality, we can impose
\begin{equation}
  \int d \eta |\hat g_1^{(i)}|^2  = \int d\eta |\hat g_2^{(i)}|^2 = 1 \; , \label{eq:normhat}
\end{equation}
where the $\eta$-space integration domain may be chosen as any convenient multiple of $2\pi$ due to the periodicity of Eqs. (\ref{eq:29sis}). Meanwhile,
\begin{equation}
  |e_1^{(i)}(\nu_i,r)|^2 + |e_2^{(i)}(\nu_i,r)|^2 = 1 \; . \label{eq:normpol}
\end{equation}
Thus, the $(e_1^{(i)}, e_2^{(i)})$ vector can be considered as definition of fluctuation polarization in the linear (Hilbert) space of Floquet solutions $[g_1^{(i)}(\eta;\nu_i,r), g_2^{(i)}(\eta;\nu_i,r)]$, corresponding to continuous spectrum fluctuations for a given $\nu_i$. Polarization is a necessary element that uniquely identifies the continuous spectrum, described by solutions of Eqs. (\ref{eq:29sis}), together with the characteristic Floquet exponent, $\nu_i(\omega, r)$, and the corresponding parallel fluctuation structures, $\hat g_1^{(i)}(\eta;\nu_i,r)$ and $\hat g_2^{(i)}(\eta;\nu_i,r)$. A pictorial representation of parallel propagation of radial singular structures of the continuous spectrum and of their polarization in the orthogonal Hilbert space is given in \figref{fig_zero}.

Here, it is also worthwhile pointing out the connection of the Floquet characteristic exponent to the generalized plasma inertia, $\Lambda$, which is an essential element of the general fishbone-like dispersion relation (GFLDR)\cite{chen16,zonca14a,zonca14b}:
\begin{equation}
i |s| \Lambda =  \delta \hat W_f  + \delta \hat W_k \;\; , \label{eq:fishlike}
\end{equation}
where $\delta \hat W_f$ and $\delta \hat W_k$ account for, respectively, the generalized potential energy due to fluid-like plasma response and the kinetic plasma behavior due to, e.g., EPs. At a given radial location, the short scale (large-$\eta$) radial structure of any considered physical fluctuation (antenna driven and/or eigenmode) satisfies the same Eqs. \eqref{eq:29sis}  describing the continuous spectrum. Thus, the large-$|\eta|$ local physical solution must be a linear superposition of the solutions of Eq. (\ref{eq:34}). Among them, generally, only two satisfy boundary conditions (outgoing/decaying wave in $\eta$-space\cite{chen16,zonca14a,zonca14b}). Renumbering them as $i=1,2$, without loss of generality, we have
\begin{equation}
  \bm x = w_1 \bm x_1 + w_2 \bm x_2 \; ,
  \label{eq:physol}
\end{equation}
where $w_1$ and $w_2$ are weights of the linear combination, which we can generally assume satisfying the normalization condition $w_1^2 + w_2^2 = 1$. The actual value of $(w_1,w_2)$ is determined consistently with the mode structure and fluctuation dispersion relation. Adopting the GFLDR theoretical framework\cite{chen16,zonca14a,zonca14b}, it is possible to show\cite{zonca14a,zonca14b}
\begin{equation}
  i \Lambda = \left.\frac{\left(i \nu_{1} P_{1}(\eta)+P_{1}^{\prime}(\eta)\right) w_{1}+\left(i \nu_{2} P_{2}(\eta)+P_{2}^{\prime}(\eta)\right) w_{2}}{P_{1}(\eta) w_{1}+P_{2}(\eta) w_{2}}\right|_{\eta=2 \ell \pi}\;\; , \label{eq:lambdanu}
\end{equation}
where $P_i(\eta)|_{\eta = 2\ell\pi} (i = 1,2)$ is referred to the $g_1$ solution of Eqs. (\ref{eq:29sis}) represented as in Eq. (\ref{eq:34}) for $\nu_i (i=1,2)$, respectively, and, for simplicity, we have assumed that the equilibrium is general but still up-down symmetric. Since the continuum fluctuation structures, Eq. (\ref{eq:eigencont}), bear the information of mode polarization, this relation shows the importance of mode polarization to assess the actual coupling of any physical fluctuation to the continuous spectrum; and further illuminates the profound connection between the structures of SAW and ISW continuous spectra and the GFLDR, and the usefulness of the present approach when extended to 3D/stellarator geometry and/or gyrokinetic descriptions. More on this important point is discussed in Section \ref{sec:org78c95f5} below.

In order to calculate $\nu_{i}$ as a function of $\omega$ for each $r$, we note that the fundamental matrix \(\mathbf{X}(\vartheta)\) of the first order system associated to Eqs. (\ref{eq:29sis}) must satisfy the following relation:
\begin{equation}
\label{eq:2pippo}
\mathbf { X } ( \vartheta + 2 \pi ) = \mathbf { X } ( \vartheta ) \mathbf { M }\;.
\end{equation}
It can be shown\cite{floquet1883equations} that \(\mathbf{M}\) does not depend on \(\vartheta\) and, therefore, is conveniently calculated for $\vartheta =0$:
\begin{equation}
\label{eq:9}
\mathbf { M } = \mathbf { X } ^ { - 1 } ( 0 ) \mathbf { X } ( 2 \pi)\;.
\end{equation}
Choosing the initial conditions such that $\mathbf { X } ( 0 ) = \mathbf { I }$, the 4x4 identity matrix in the present case, we obtain:
\begin{equation}
\label{eq:11}
\mathbf { M } = \mathbf { X } ( 2 \pi )\;.
\end{equation}
The eigenvalues of \(\mathbf{M}\) are the characteristic multipliers of the system and can be expressed in terms of the characteristic Floquet exponents by the following relation:
\begin{equation}
\label{eq:12}
\rho _ { i } = e ^ {i  2 \pi \nu _ { i }  }\;.
\end{equation}
Corresponding eigenvectors can be used to calculate parallel mode structures and polarization vectors, Eq. (\ref{eq:eigencont}); e.g., to  calculate $P_i'(2\pi)/P_i(2\pi)$ and the value of the generalized inertia from Eq. (\ref{eq:lambdanu}). The same information can be adopted as boundary conditions for the calculation of fluid and kinetic potential energies in the GFLDR\cite{chen16,zonca14a,zonca14b}. However, doing so is outside the scope of the present work.
\subsubsection{Hill's equation}
\label{sec:periodichil}
In the decoupled case ($\kappa_g = 0$), Eqs. (\ref{eq:29sis}) become a system of two independent differential equations in the form: 
\begin{equation}
\label{eq:2}
\frac { d ^ { 2 } x } { d \eta ^ { 2 } } + V ( \eta; \omega )  x = 0\;,
\end{equation}
where $x=g_{1},g_{2}$; which are a particular cases of Hill's equation\cite{hill1886part,magnus2013hill,Denk1995}. Note that, consistent with previous notations, parametric dependences on $r$ are left implicit for notation simplicity, while the dependence on $\omega$ of the potential function $V(\eta; \omega)$ is indicated explicitly. Reducing each one of these equations to a first order system and using the results summarized in the previous subsection, we know that they will have Floquet solutions satisfying the following property:
\begin{equation}
\label{eq:4}
\boldsymbol x_{i} ( \eta + 2 \pi; \omega ) = \rho_{i} \boldsymbol x_{i} ( \eta ;\omega )\;,
\end{equation}
where $i=1,2$. We can show that  the following relation holds\cite{abramowitz1965handbook}:
\begin{equation}
\label{eq:5pippo}
\rho_{i} ^ { 2 } - 2 A ( \omega ) \rho_{i} + 1 = 0\;,
\end{equation}
with:
\begin{equation}
\label{eq:6}
A ( \omega ) = \frac { 1 } { 2 } \left(\boldsymbol x _ {1 0 } ( 2 \pi ; \omega ) + \boldsymbol x_ { 01 } ^ { \prime } ( 2 \pi ; \omega ) \right)\;,
\end{equation}
where \(\boldsymbol x_{10}\) and \(\boldsymbol x_{01}\) are the particular solutions of the first order system associated to each Hill's equation such that \(\boldsymbol x_{10}(0) = (1,0)\) and \(\boldsymbol x_{01}(0) = (0,1)\). In the particular case of even $V(\eta; \omega)$, it can be shown that \(A(\omega) = x_{10}(2\pi;\omega)\). Solving Eq. (\ref{eq:5}) gives the following expression for the multipliers:
\begin{equation}
\label{eq:16}
\rho _ { 1,2 } = \frac { + A \pm \sqrt { A ^ { 2 } - 4 } } { 2 }\;.
\end{equation}
If \(|A^{2}|<2\), roots are complex conjugates with unity absolute value while, if \(|A^{2} = 2|\), they coincide and are equal to \(1\). Finally, if \(|A^{2} > 2|\), roots are real and reciprocal to each other. In particular, the first case describes the frequency continuum while the third the frequency gaps. From this expression is possible to calculate the characteristic exponent from the value of \(A(\omega)\). In the first case we obtain:
\begin{equation}
\label{eq:19}
\cos{(\nu 2 \pi)} = \frac{A(\omega)}{2}\;,
\end{equation}
where $\nu$ is real. From this analysis we expect up to two (opposite) values of $\nu$ for each $\omega$ in the decoupled case, corresponding to waves
propagating in opposite directions along the field lines. This result can be generalized to higher dimensionality, e.g. see Ref.\citenum{Denk1995}, to describe the coupled system, where we expect up to two (pairs of opposite) $\nu$ values for each $\omega$ corresponding, again, to waves propagating in opposite directions along the field lines.
\section{Numerical results}
\label{sec:org78c95f5}
\begin{figure}[!htpb]
\centering
\includegraphics[width=0.8\textwidth]{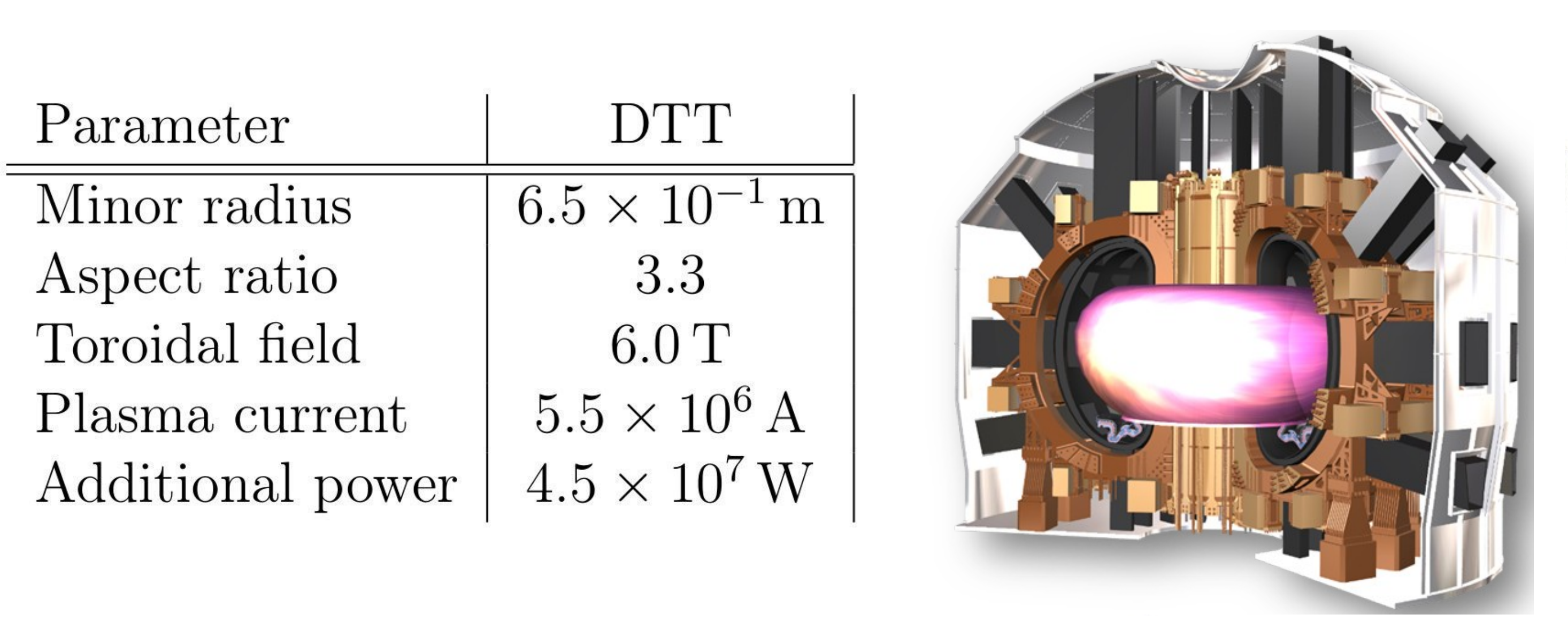}
\caption{DTT main parameters \cite{albanese19}.\label{fig_DTT}}
\end{figure}

In this Section, in order to show the generality of the present approach, we calculate the frequency continuum of SAW and ISW waves in a Divertor Tokamak Test Facility (DTT) reference scenario. DTT is an Italian project \cite{albanese17,albanese19} sponsored by the EUROfusion consortium with the goal of designing a new machine capable of eventually integrating all relevant physics and technological issues concerning power exhaust solutions for DEMO. The main DTT parameter are reported in \figref{fig_DTT}. In order to isolate the role of various equilibrium geometry effects in SAW-SAW, ISW-ISW and SAW-ISW couplings and discuss the corresponding structures in the continuous spectra, we illustrate first the decoupled SAW and ISW continuous spectra by artificially letting $\kappa_g=0$ (Section \ref{sec:sawdec}). The realistic situation with coupled SAW and ISW spectra is then addressed in Section \ref{sec:orge42a33e}. Numerical results obtained with the present approach are compared with those by the MARS code adopting the conventional method, showing excellent agreement.
\subsection{Normalized equations}
\label{sec:org65c228a}
The set of equations describing SAWs coupled with ISW can be re-written in terms of the following dimensionless quantities:
\begin{equation}
\label{eq:15}
\Omega = \frac { \omega R_ { 0 } } { \overline { v } _ { A 0 } }, \quad
\hat{J}_{\eta}^{2} \equiv \frac{J_{\eta}^{2}\bar{B}_{0}^{2}}{R_{0}^{2}}, \quad \hat{\rho}_{m0}=\frac{\rho_{m0}}{\bar{\rho}_{m0}}, \quad \hat{B}_{0}=\frac{B_{0}}{\bar{B}_{0}}, \quad \hat{\kappa}_{g}= \kappa_{g}R_{0}\;,
\end{equation}
where \(\overline { v } _ { A 0 }\) is the Alfv\'en velocity on the magnetic axis. The non-dimensional form of Eqs. \eqref{eq:29sis} then reads now 
\begin{subequations}\label{sysdmls}
\begin{align}
\label{eq:36}
\left( \partial _ { \eta } ^ { 2 } + \hat { J } _ { \eta } ^ { 2 } \hat{\rho}_{m0} \Omega ^ { 2 } \right) g _ { 1 } &=  2  \hat{B} _ { 0 }   \hat{\rho}_{m0}^ { 1 / 2 } \hat { J } _ { \eta } ^ { 2 } \hat{\kappa} _ { g } \Omega\, g _ { 2 }\;, \\
\left( \partial _ { \eta } ^ { 2 } - |\bm \nabla r | \partial _ { \eta } ^ { 2 } | \bm\nabla r | ^ { - 1 } + \frac { 2 } { \Gamma \beta } \hat{\rho}_{m0} \hat { J } _ { \eta } ^ { 2 } \Omega ^ { 2 } \right) g _ { 2 } &=  2  \hat{B} _ { 0 }  \hat{\rho}_{m0}^ { 1 / 2 } \hat { J } _ { \eta } ^ { 2 } \hat{\kappa} _ { g } \Omega\, g _ { 1 } \;.\label{condhgvat}
\end{align}
\end{subequations}
Eqs. \eqref{sysdmls} are studied numerically in the remaining part of this work for the DTT reference scenario illustrated in the following subsection.
\subsection{DTT reference scenario}
\label{sec:referencescenario}

\begin{figure}[!htpb]
\centering
\includegraphics[width=0.49\textwidth]{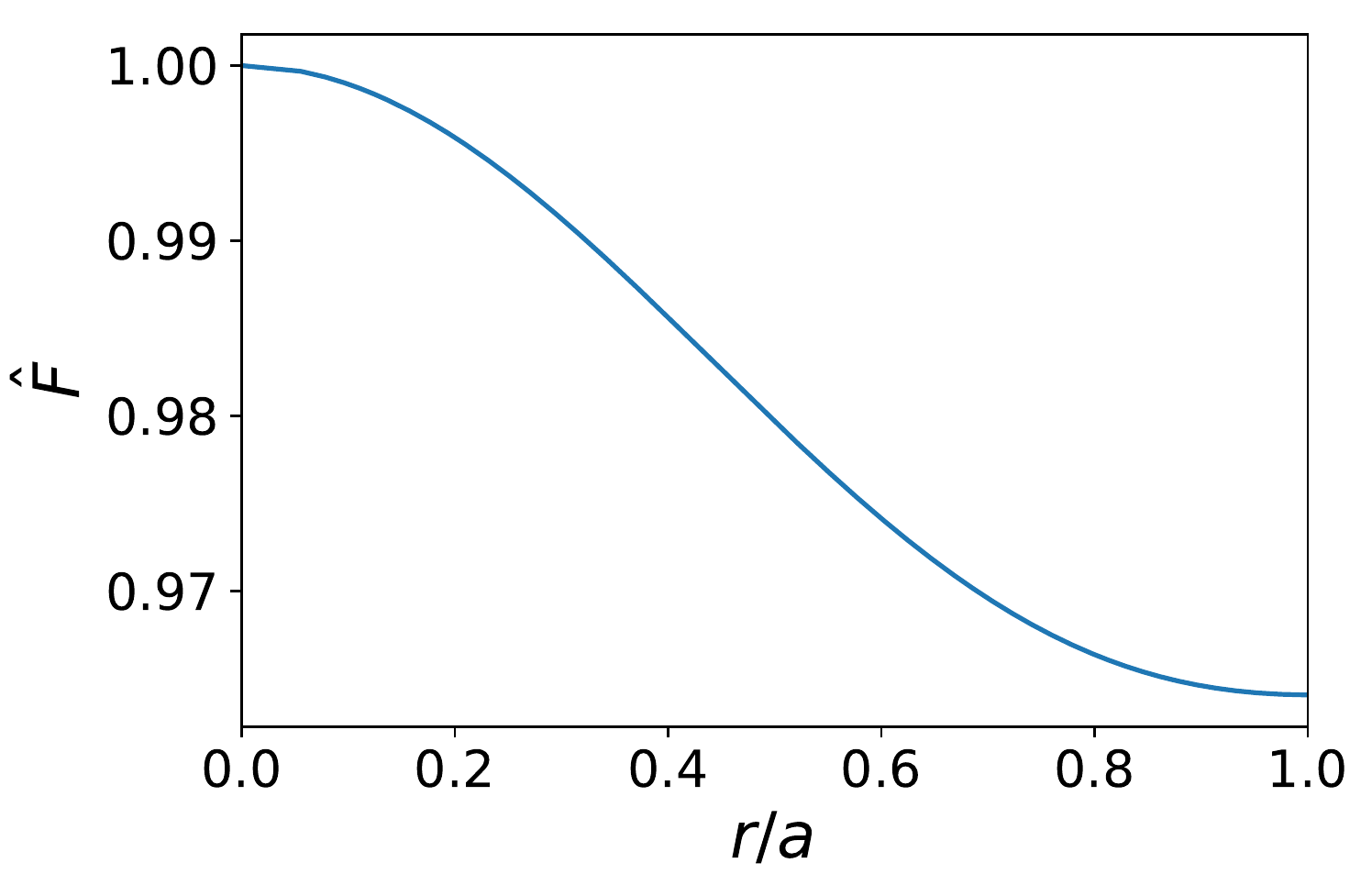}
\includegraphics[width=0.43\textwidth]{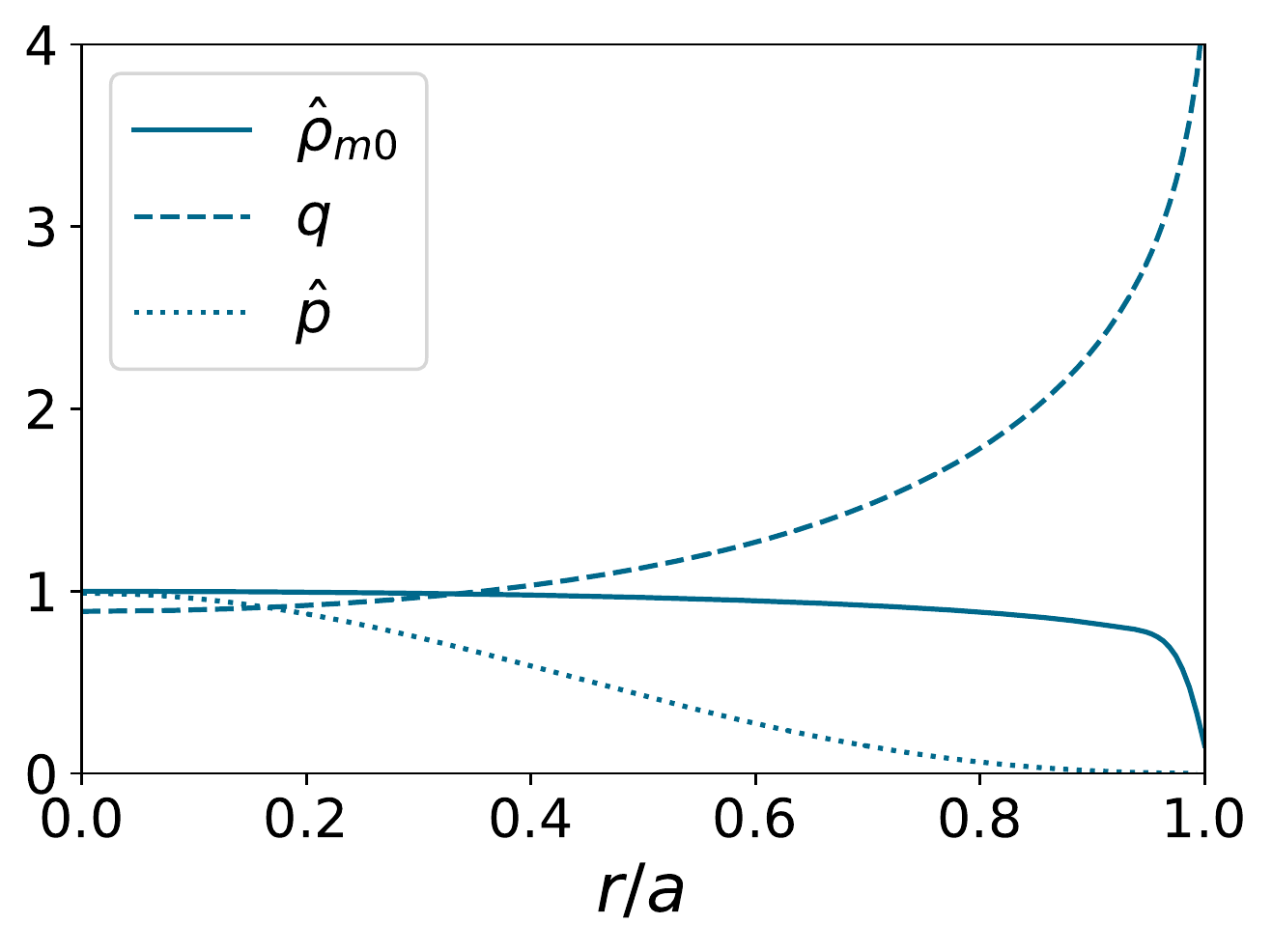}
\caption{Plots of the main profiles for the reference double null scenario as a function of $r/a$. Every quantity except $q$ is normalized to its value on the magnetic axis:  flux function (left-hand panel), kinetic pressure, density together with the safety factor $q$ (right-hand panel). \label{fig_profiles}}
\end{figure}

\begin{figure}[!htpb]
\centering
\includegraphics[width=0.4\textwidth]{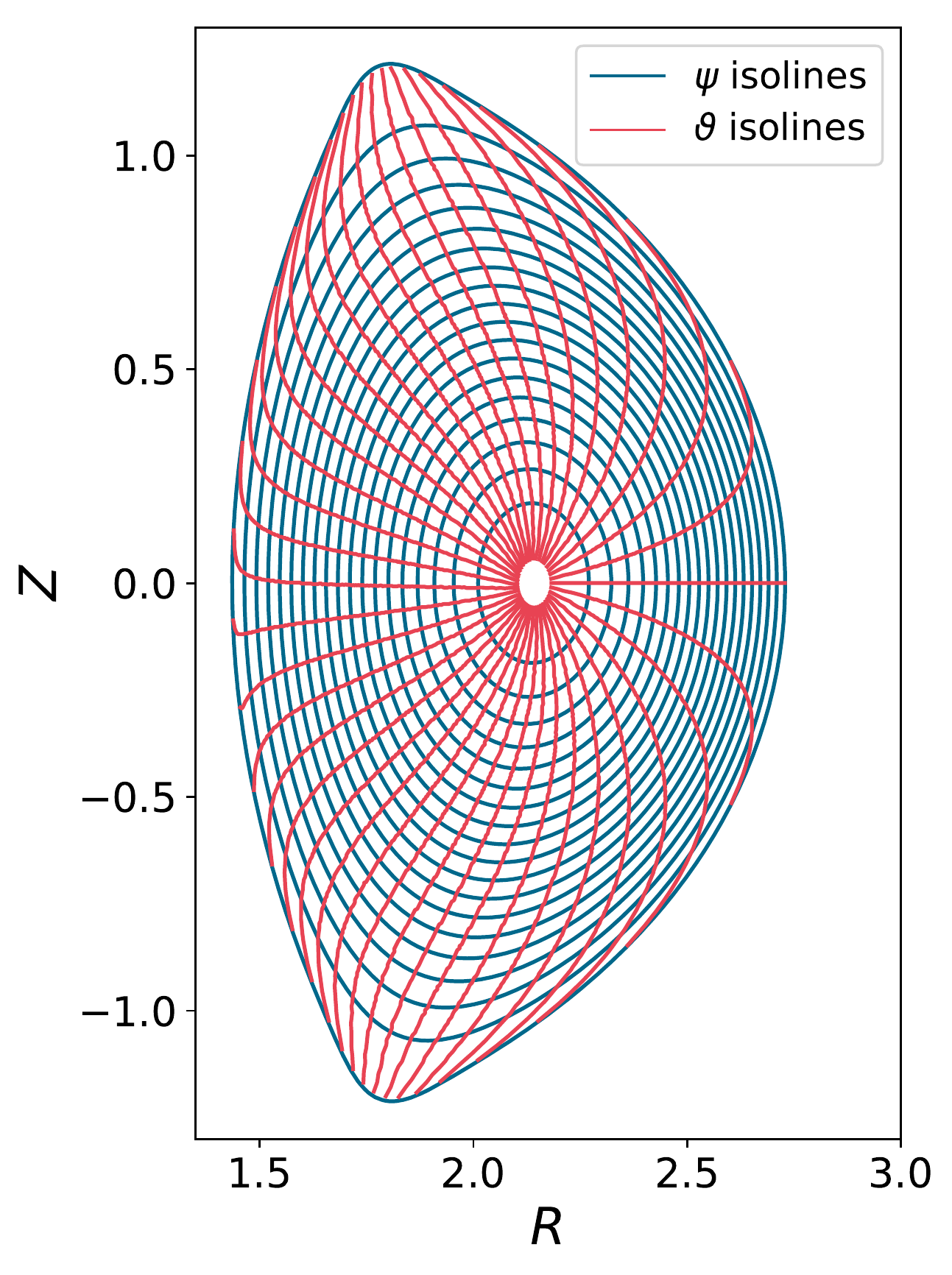}
\includegraphics[width=0.4\textwidth]{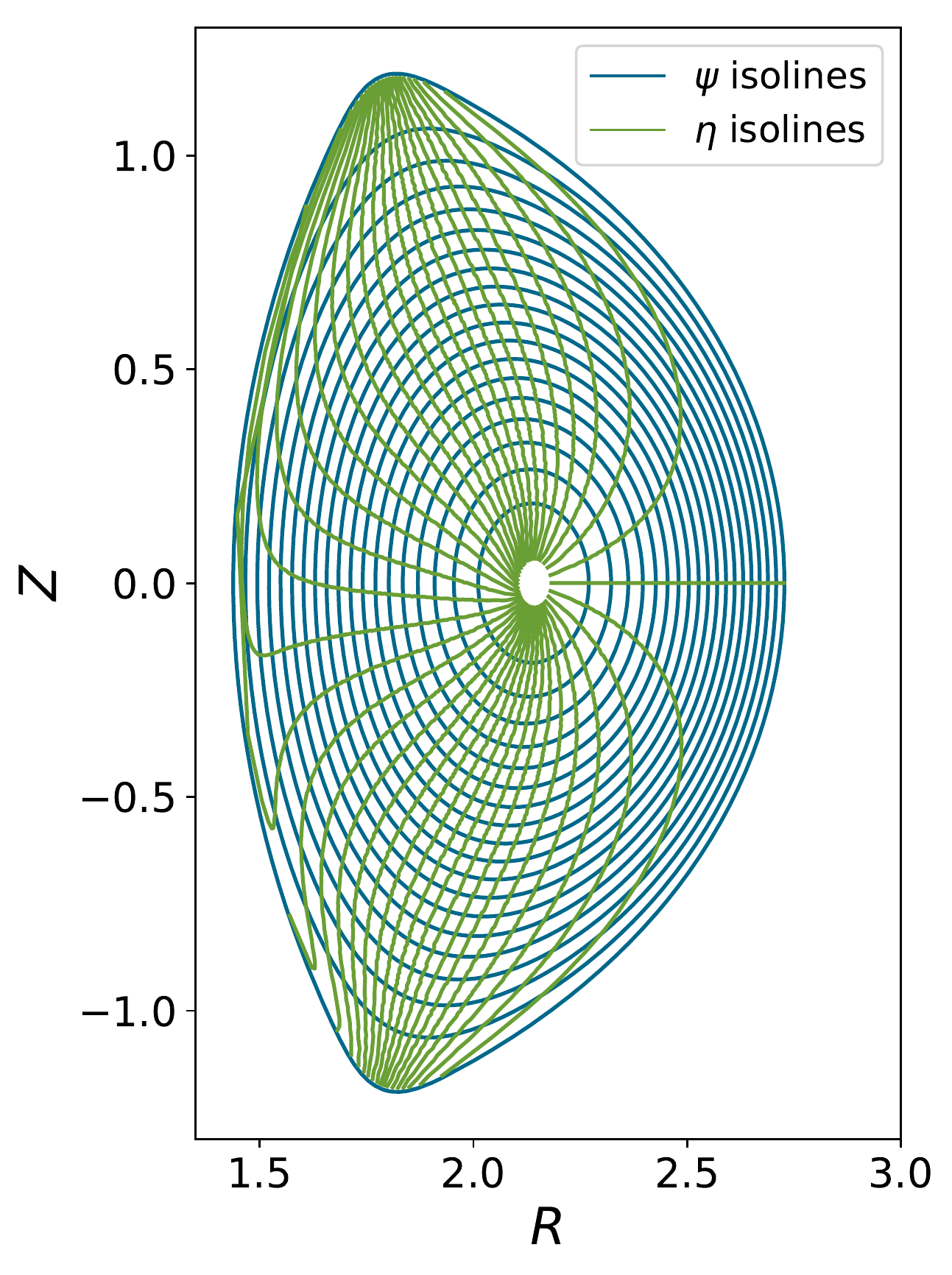}
\caption{(Color online) Contour lines of $\psi$ and isolines of the $\vartheta$ coordinates (left-hand panel, red lines) and $\eta$ coordinates (right-hand panel, green lines). $R$ and $Z$ are expressed in meters.\label{fig_isolines}}
\end{figure}

The magnetic equilibrium has been originally calculated by means of the free bounddary equilibrium evolution code CREATE-NL \cite{CREATE-NL} and further refined and mapped on flux coordinates by using the high-resolution equilibrium solver CHEASE \cite{CHEASE}. We consider a double null setting whose basic profiles are depicted in \figref{fig_profiles} as a function of the normalized toroidal radius $r/a$. We note that, for the analyzed case, the kinetic pressure on axis is $1.0768\times10^{6}$Pa, the flux function $F$, see Eq.(\ref{eq:27}), on axis is $12.45$Vs, while the density is $10^{19}$m$^{-3}$. 
Moreover, the normalized density profile has been obtained by studies of plasma scenario formation using the fast transport simulation code METIS \cite{METIS}. For the addressed scenario, the $(\psi,\vartheta)$ and $(\psi,\eta)$ isosurfaces are depicted in \figref{fig_isolines}.
\subsection{SAW continuous spectrum}
\label{sec:sawdec}
\begin{figure}[!htpb]
\centering
\includegraphics[width=0.6\textwidth]{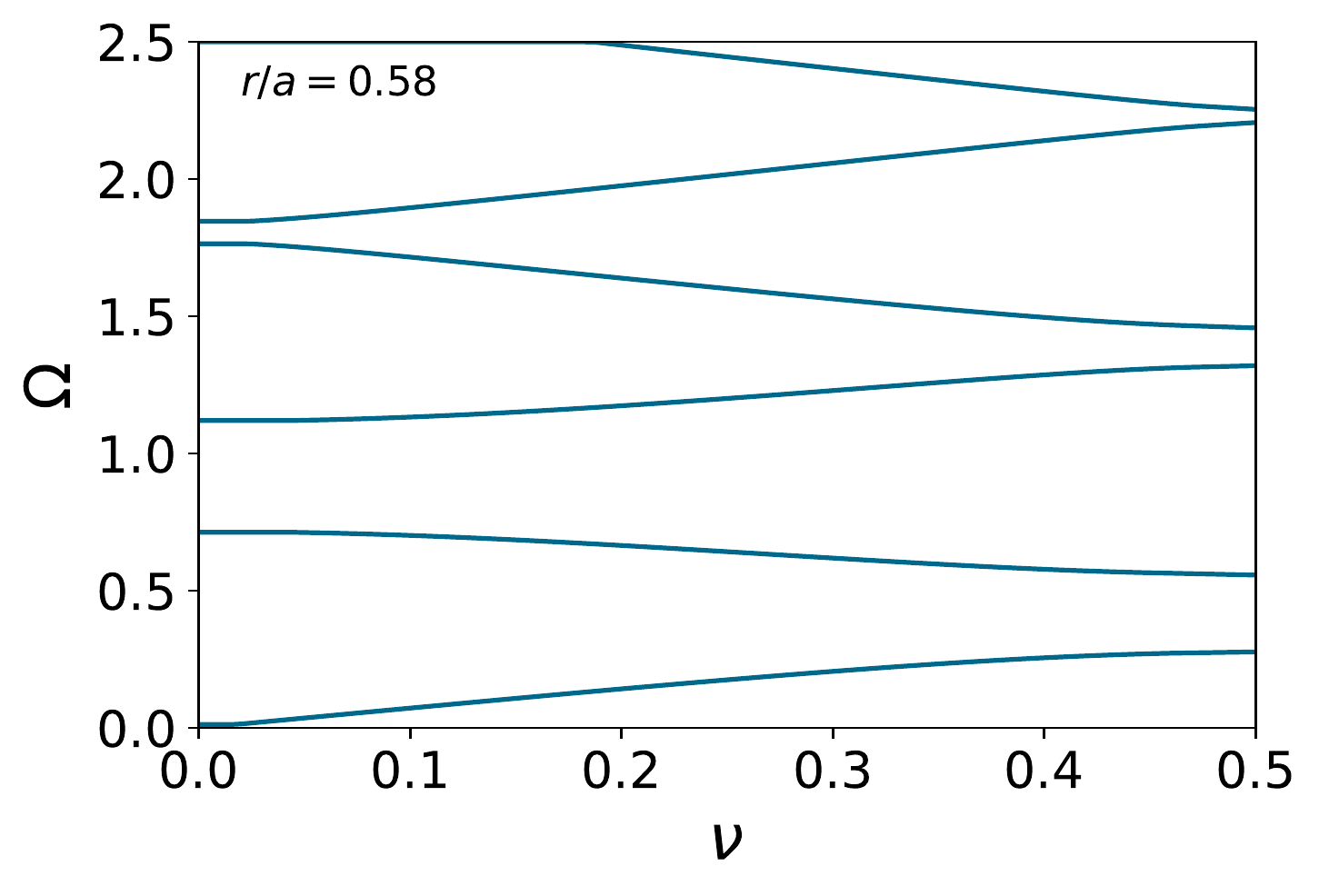}
\caption{Plot of the local dispersion curves $\nu(\Omega)$  for fixed $r/a=0.58$.}
\label{fig:sawsinglesurf}
\end{figure}

\begin{figure}[!htpb]
\centering
\includegraphics[width=0.6\textwidth]{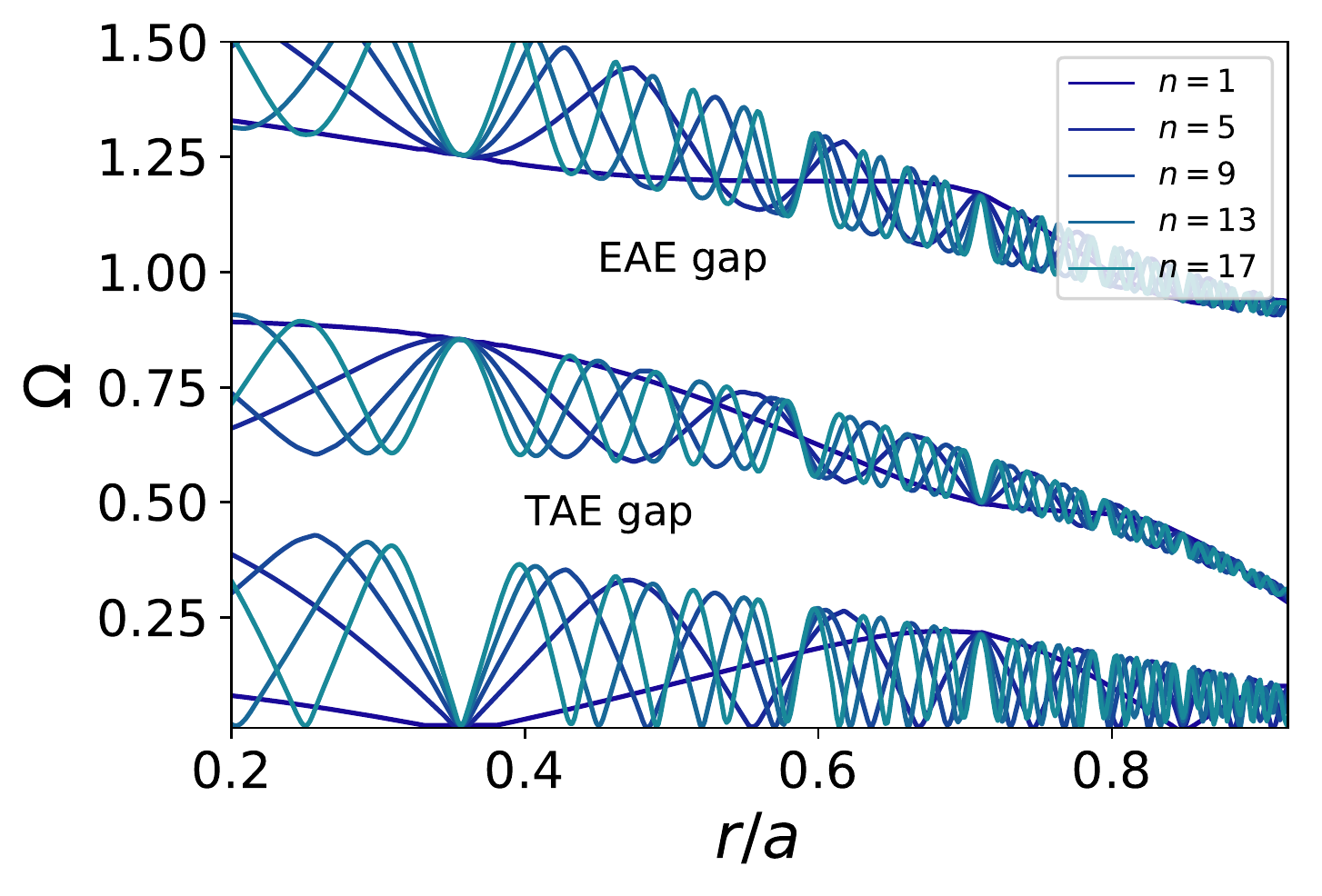}
\caption{(Color online) SAW continuous spectrum as a function of the normalized radial position for several values of the toroidal mode number as indicated in the plot.}
\label{fig:sawmultisurf}
\end{figure}

The equation describing SAW continuous spectrum can be obtained from Eqs. (\ref{sysdmls}) by neglecting the coupling terms with the ISW branch. Thus, we have
\begin{equation}
\label{eq:25}
\left( \partial_{\eta}^{2}+ \hat{J}_{\eta}^{2} \hat{\rho}_{m0} \Omega^{2}\right)g_{1}=0\;.
\end{equation}
In order to illustrate with a practical application the procedure introduced in Sec. \ref{sec:periodicflo}, we first show the local dispersion curves $\nu(\Omega)$ obtained integrating Eq.(\ref{eq:25}) at fixed $r/a$, \idest on a given $\psi$ isosurface, for different $\Omega$ values. These curves are plotted in \figref{fig:sawsinglesurf} for the reference value $r/a=0.58$, where, for convenience, $\nu$ is chosen as abscissa showing different branches and gaps.
Noting Eq. \eqref{eq:deco} and assuming a Fourier decomposition of the fluctuations mapped back from ballooning to real space\cite{lu12}, we can write
\begin{equation}
\hat J_\eta \hat B_0 R_0 k_{\parallel m, n} = nq - m \; . \label{eq:bragg1}
\end{equation}
Thus, two counter-propagating SAWs can form a standing wave and cause the  formation of a frequency gap in the continuous spectrum\cite{chance77,cheng85,Zonca1992prl,Zonca1993fluids}, when the Bragg reflection condition is satisfied; that is:
\begin{equation}
  k_{\parallel m+p, n} = - k_{\parallel m, n}\;\;\;\;\; \Rightarrow \;\;\;\;\;  k_{\parallel m, n} = \frac{p}{2\hat J_\eta \hat B_0 R_0} \; , \;\; p \in \mathbb Z \; . \label{eq:bragg2}
\end{equation}
Due to the periodicity of Floquet solutions, Eq. (\ref{eq:34}), Eq. (\ref{eq:33}) can be effectively reduced to the first Brillouin zone, $0\leq \nu\leq 1/2$, and frequency gaps are expected to appear at\cite{zonca14a,zonca14b} (cf. Appendix \ref{app:A}) 
\begin{equation}
  \Omega^2 =  \frac{p^2/4}{\hat \rho_{m0} \left\langle \hat J_\eta^2 \right\rangle_{g_1}} \; , \;\; p \in \mathbb Z\; ; \label{eq:SAWgap}
\end{equation}
with \begin{equation}
\left\langle \hat J_\eta^2 \right\rangle_{g_1} \equiv \frac{\int d \eta \hat J_\eta^2 |g_1|^2 }{\int d \eta |g_1|^2 } \; .
\end{equation}
consistent with Eqs. (\ref{eq:bragg1}) and (\ref{eq:bragg2}). As anticipated above, dispersion curves $\nu = \nu (\Omega,r/a)$, \idest the dimensionless form of Eq. (\ref{eq:356}), include all the relevant information for the construction of continuous spectra. Using Eq.(\ref{eq:34}) and combining the results of different flux surfaces, we obtain the SAW continuous spectrum as a function of $r/a$ as depicted in \figref{fig:sawmultisurf}, where continua for various toroidal mode numbers are shown in different colors and illustrate the formation of toroidicity- as well as ellipticity-induced frequency gaps at, respectively, $\Omega \simeq  1/2$ (TAE\cite{cheng86}) and $\Omega \simeq  1$ (EAE\cite{Betti1991,Betti1992}). Higher frequency gaps could also be readily plotted by means of the dispersion curves reported in \figref{fig:sawsinglesurf} but are omitted here to illustrate results more clearly.

\begin{figure}[!htpb]
\centering
\includegraphics[width=0.6\textwidth]{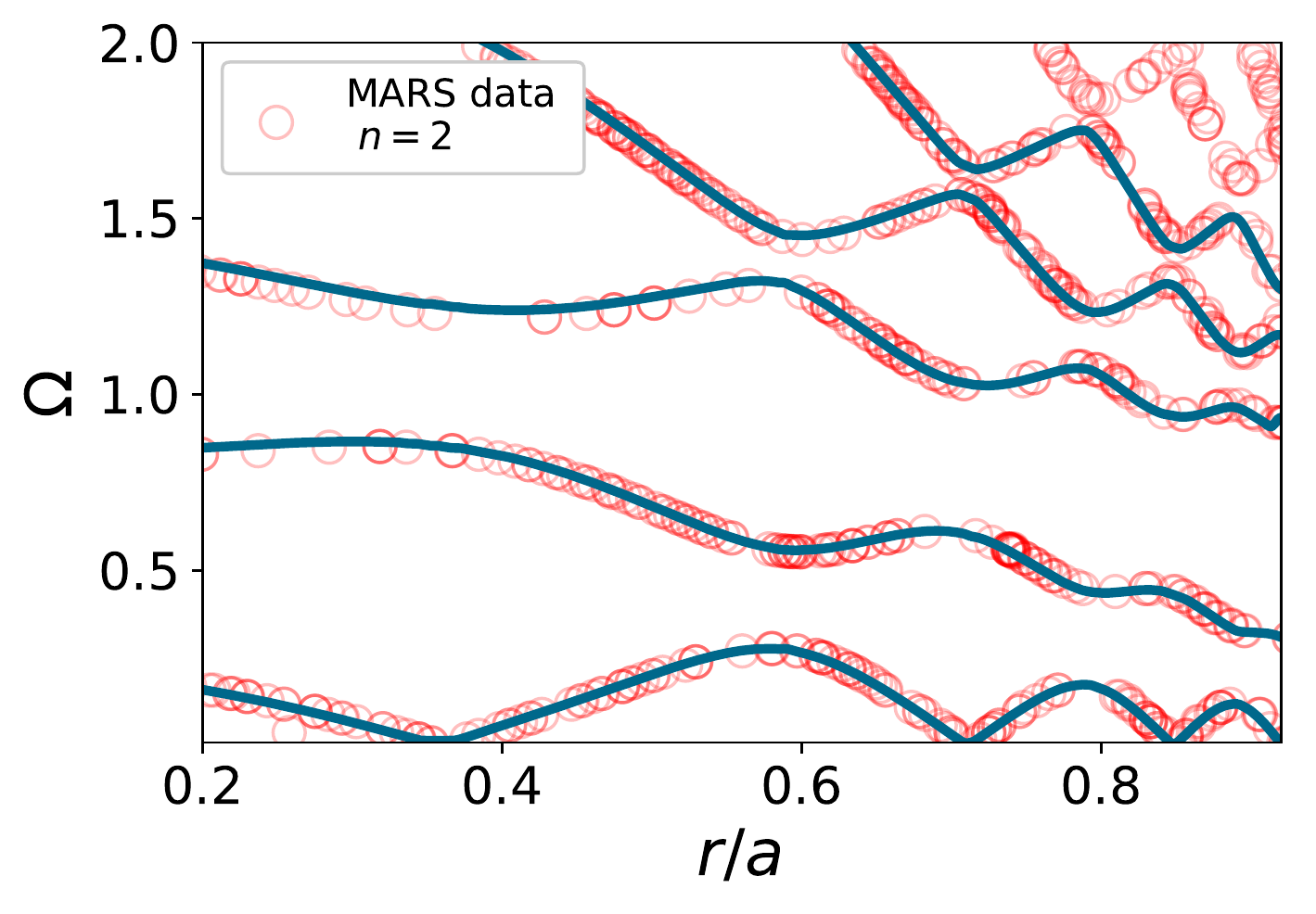}
\caption{(Color online) Comparison of the decoupled SAW continuous spectrum calculated within the proposed approach (solid lines) and the results obtained by means of the MARS code (red bullets), in the case $n=2$.}
\label{fig:sawcomparison}
\end{figure}

We now compare the results with the spectrum calculated by MARS code\cite{Bondeson1992}. The continuous spectrum is obtained assigning a guess real frequency $\omega_{c} = \omega_{\mathrm{guess}}$ and looking for converged solutions at $\omega = \omega_{\mathrm{c}}$, which correspond to a perturbed velocity whose toroidal contravariant $(m,n)$ Fourier components $v^{\phi}_{m,n}(r)$ exhibit a divergent radial dependence of the type $v^{\phi}_{m,n}(r) \propto 1/(r-r_{\mathrm{c}})$. When this condition is recognized, the points $(r_{\mathrm{c}}/a,\omega_{\mathrm{c}})$ in the plane $(r/a,\omega)$ are considered as the location of the continuous spectrum. The comparison for the toroidal mode number $n=2$ is depicted in \figref{fig:sawcomparison}, showing good agreement between the two approaches. 

As anticipated above, one of the peculiar features of the approach proposed by the present work is that the dispersion curves are independent of $n$ and, therefore, the evaluation of the continuous spectrum for different toroidal mode numbers merely requires the numerical solution of an algebraic equation, Eq. (\ref{eq:356}). Another interesting property of the present methodology is the freedom in the choice of the radial mesh due to the (radial) local nature of the problem that we are solving. This feature can be exploited to more accurately describe regions where the continuous spectrum is more convoluted. Finally, the present approach is readily extended to 3D geometries and kinetic analyses, as noted in the Introduction and Section \ref{sec:org90d93a1}.
 
The same analysis can be repeated for the ISW continuum. In fact, Eq. (\ref{condhgvat}) can be cast in the same form as Eq. \ref{eq:25} using Boozer coordinates and normalizing the frequency to the local sound speed rather than the Alfvén speed on axis. Results, thus, would be exactly the same as in \figref{fig:sawmultisurf} upon a suitable rescaling of ISW continuum frequencies. In particular, a toroidicity induced gap would appear for $p=1$ in Eq. (\ref{eq:SAWgap}); an ellipticity induced gap would occur for $p=2$ and so on. These frequency gaps have not received significant attention in the literature since, as already pointed out, the ISW branch is typically affected by strong Landau damping\cite{chavdarovski09,zonca10,chavdarovski14,liu17,zhang16,Bierwage2017,Lauber2013} and, therefore, discrete modes that are possibly located in these gaps are characterized by high excitation thresholds\cite{chen16,zonca14a,zonca14b,chen17}. Meanwhile, the features of ISW-ISW and SAW-SAW couplings due to equilibrium geometry are identical, as are the corresponding structures of continuous spectra. Therefore, a detailed calculation of the uncoupled ISW continuum is omitted here.

\subsection{Coupled system}
\label{sec:orge42a33e}
\begin{figure}[!htpb]
\centering
\includegraphics[width=0.6\textwidth]{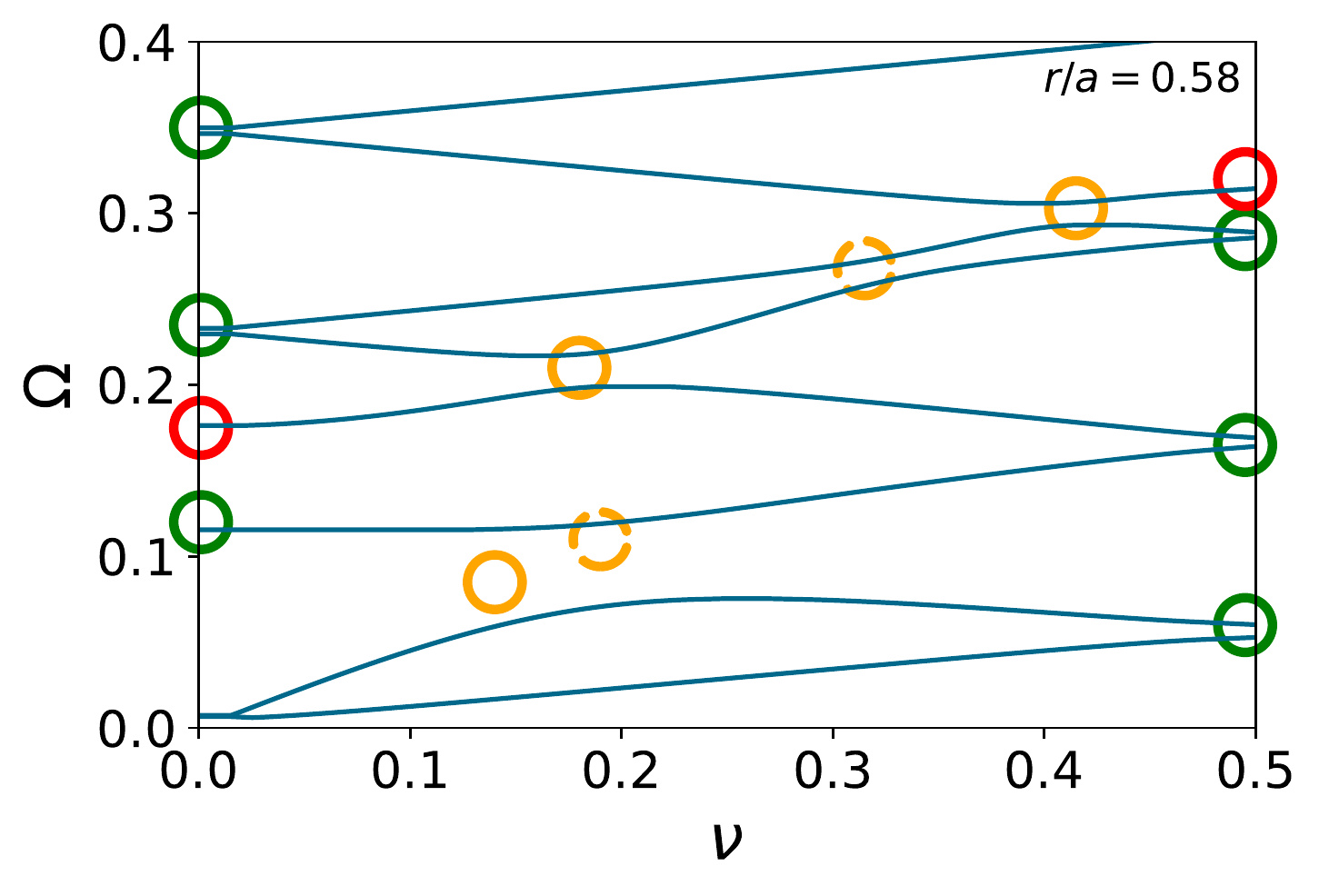}
\caption{Local dispersion curves $\nu(\Omega)$ for the SAW-ISW coupled continuum at $r/a=0.58$. Colored circles denote the interaction of ISW-ISW (green), SAW-SAW (red) and SAW-ISW (orange). Solid circles corresponding to interactions between counter-propagating waves, while dashed circles to interactions between co-propagating waves.}
\label{fig:coupsinglesurf}
\end{figure}

In the general case, we cannot neglect the SAW-ISW coupling due to equilibrium non-uniformities and we need to study the coupled system of Eqs. (\ref{eq:36}) and (\ref{condhgvat}).
In order to calculate the structures of SAW and ISW continuous spectra, we apply the procedure described in Section \ref{sec:org90d93a1}. For the reference value $r/a=0.58$, the result is depicted in \figref{fig:coupsinglesurf}, reproducing the expected behavior.
A change of the topology emerges with respect to the uncoupled case in the vicinity of the intersection points between SAW and ISW dispersion curves. In particular, it is worthwhile noting that standing waves and corresponding frequency gap formation can occur not only at the Bragg reflection condition, Eq. (\ref{eq:bragg2}), due to the interaction of two counter-propagating SAWs or ISWs (cf. green and red solid circles in \figref{fig:sawsinglesurf}), but also when
\begin{equation}
  k_{\parallel m+p, n} \left( \frac{\Gamma \hat \beta}{2} \right)^{1/2}= - k_{\parallel m, n}\;\;\;\;\; \Rightarrow \;\;\;\;\;  k_{\parallel m, n} = \frac{p}{\hat J_\eta \hat B_0 R_0} \frac{(\Gamma \hat \beta/2)^{1/2}}{1+(\Gamma \hat \beta/2)^{1/2}}\; , \;\; p \in \mathbb Z \; ,\label{eq:bragg3}
\end{equation}
where $\hat \beta \simeq \bar \beta$ is rigorously defined in Eq. (\ref{eq:hatbeta}).
In \figref{fig:sawsinglesurf}, regions where Eq. (\ref{eq:bragg3}) is satisfied are denoted by the full orange circles, corresponding to a forward propagating ISW, modulated by the equilibrium non-uniformity within the magnetic flux surface, forming a standing wave by interaction with a backward scattered SAW. Modulated forward propagating ISW can also interact with a forward propagating SAW, when
\begin{equation}
k_{\parallel m-p, n} \left( \frac{\Gamma \hat \beta}{2} \right)^{1/2}= k_{\parallel m, n}\;\;\;\;\; \Rightarrow \;\;\;\;\;  k_{\parallel m, n} = \frac{p}{\hat J_\eta \hat B_0 R_0} \frac{(\Gamma \hat \beta/2)^{1/2}}{1-(\Gamma \hat \beta/2)^{1/2}}\; , \;\; p \in \mathbb Z \; .\label{eq:bragg4}
\end{equation}
In this case, however, a standing wave cannot be formed, as it is clearly shown by the dashed orange circles in \figref{fig:sawsinglesurf}. The gap formation at low frequency for $0.1<\nu<0.2$ is due to the interaction denoted by the full orange circle and satisfying Eq. (\ref{eq:bragg3}) for $p=1$.

\begin{figure}[!htpb]
\centering
\includegraphics[width=0.6\textwidth]{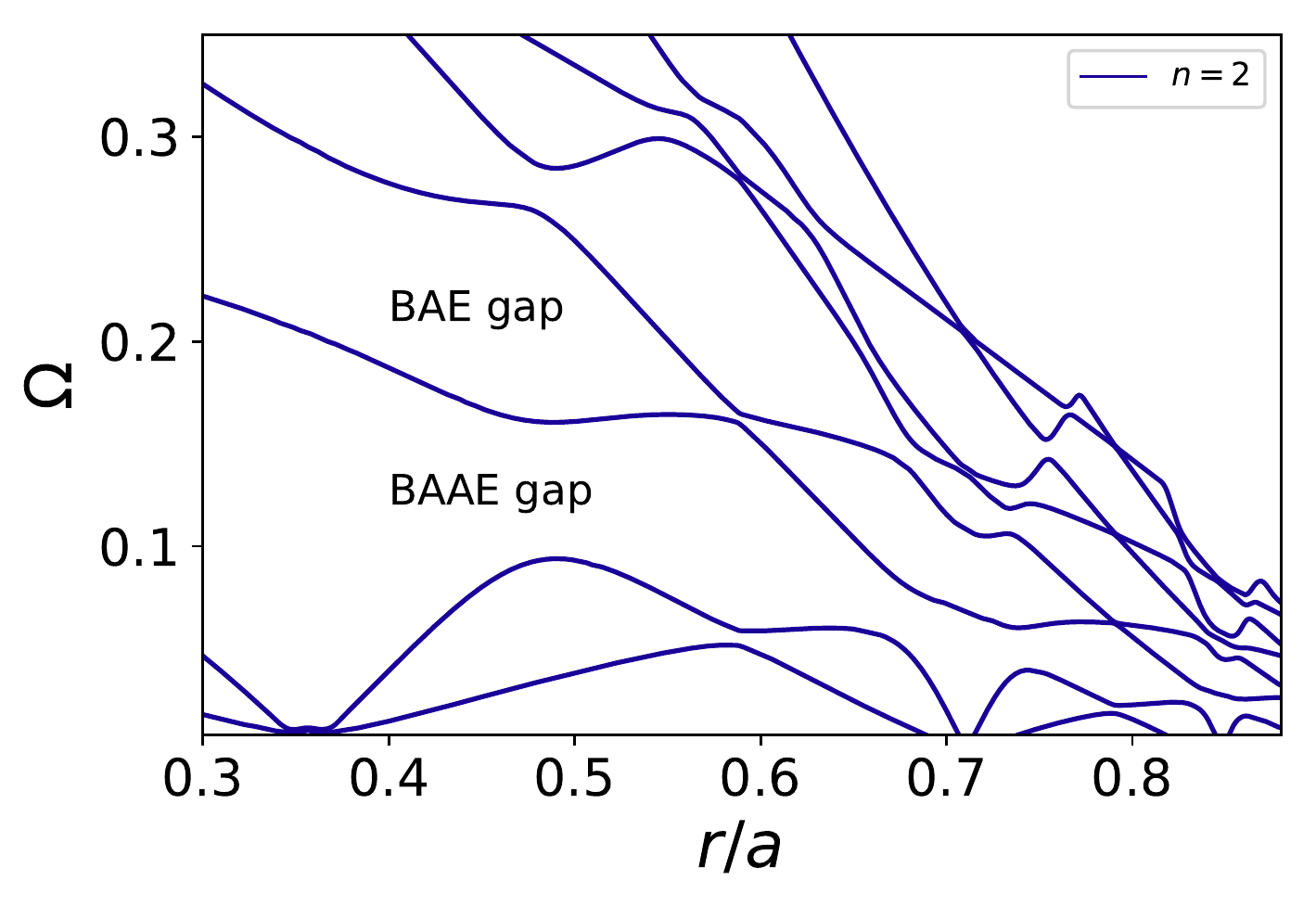}
\caption{(Color online) SAW-ISW coupled continuum as a function of $r/a$ for $n=2$.}
\label{fig:coupmultisurfn2}
\end{figure}

\begin{figure}[!htpb]
\centering
\includegraphics[width=0.6\textwidth]{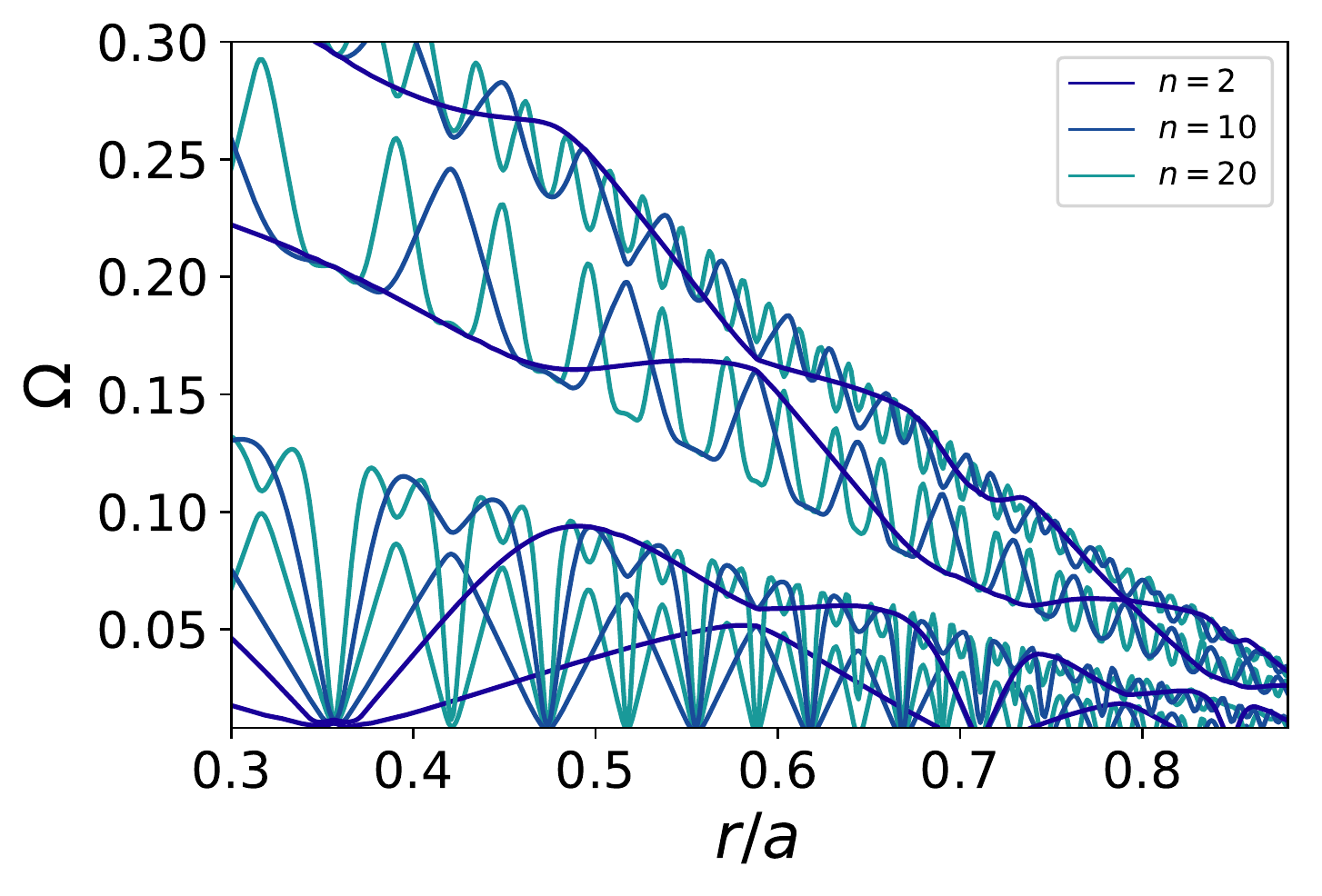}
\caption{(Color online) SAW-ISW coupled continuum as a function of $r/a$ for three toroidal wave numbers as indicated in the plot. For the sake of clarity we plot only the first four branches.}
\label{fig:coupmultisurf}
\end{figure}

\begin{figure}[!htpb]
\centering
\includegraphics[width=0.6\textwidth]{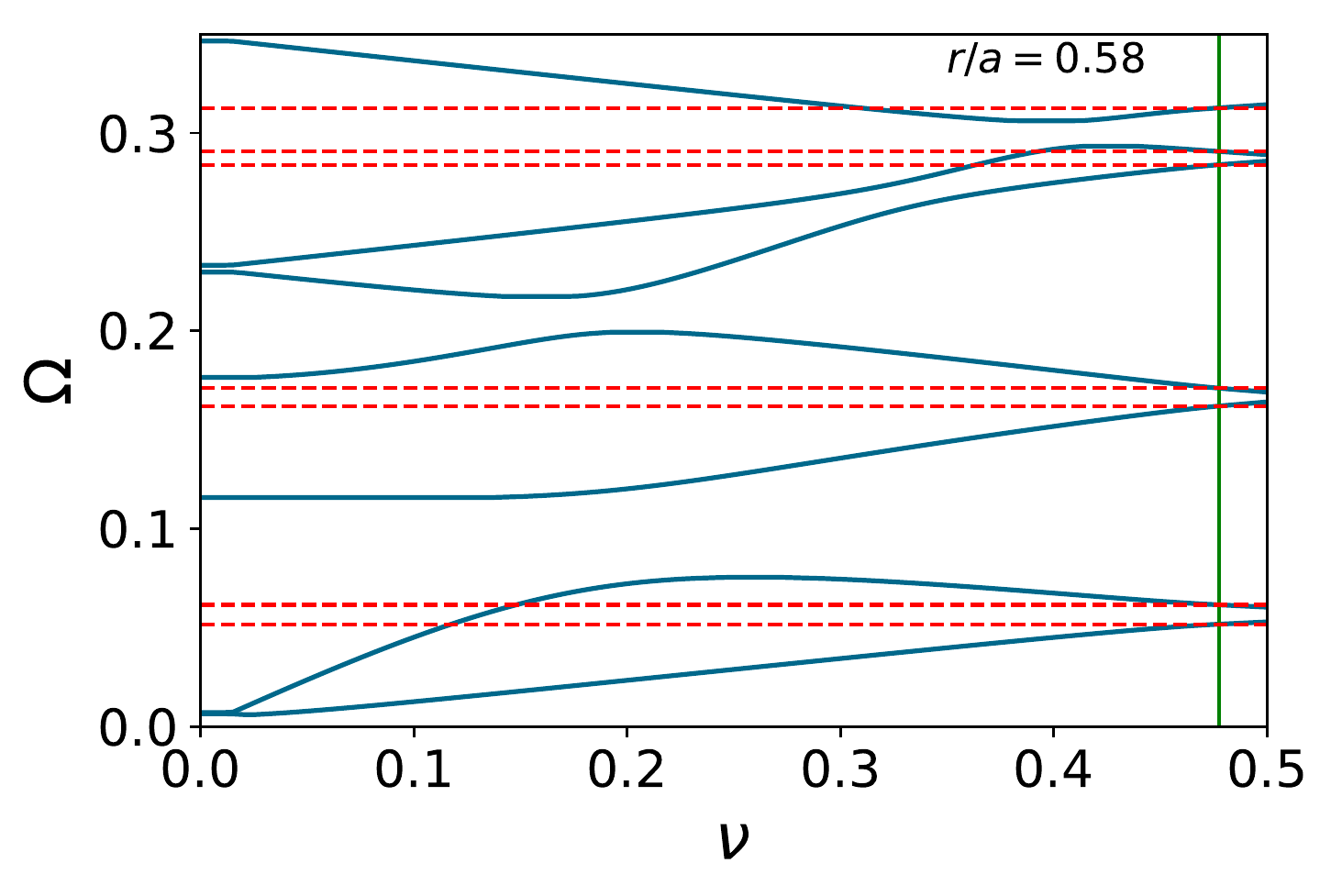}
\caption{Local dispersion curves $\nu(\Omega)$ for the SAW-ISW coupled continuum at $r/a=0.58$. The green vertical line indicates $\nu = n q(r) \bmod 1$, with $n=2$. Continuum frequencies are marked with red dashed lines.}
\label{fig:alfvenicity}
\end{figure}

\begin{figure}[!htpb]
\centering
\includegraphics[width=0.6\textwidth]{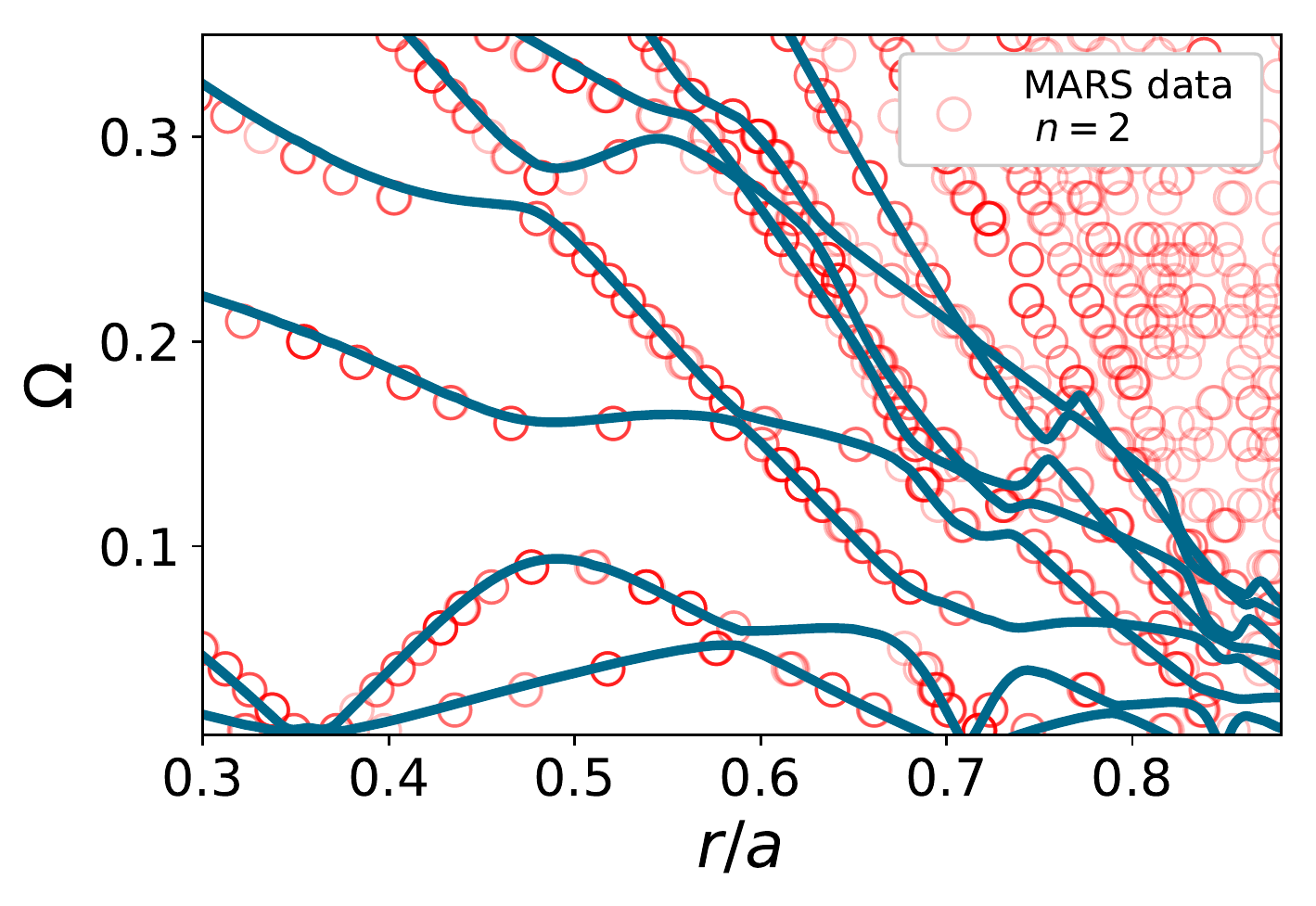}
\caption{(Color online) Comparison of the outputs of MARS code (red bullet) and the SAW-ISW coupled continuum (solid lines), for $n=2$.}
\label{fig:coupcomparison}
\end{figure}

Combining the results of different flux surfaces, we finally get the behavior of the continuous spectra as shown in \figref{fig:coupmultisurfn2} for $n=2$ and in \figref{fig:coupmultisurf} for different toroidal mode numbers.
The richness of the continuous spectrum at low frequency\cite{chu92,huysmans95,goedbloed98,vanderholst00} is readily noted, with the frequency gaps due to SAW-ISW and/or ISW-ISW couplings, dubbed Beta induced Alfvén Acoustic Eigenmode (BAAE) gap\cite{gorelenkov09,gorelenkov07a,gorelenkov07b}, as well as SAW-ISW and/or SAW-SAW couplings, dubbed Beta induced Alfvén Eigenmode (BAE) gap\cite{chu92,Heidbrink1993,Turnbull1993}, shown in \figref{fig:coupmultisurf}. The complex structures of the continuous spectrum are further illustrated by adding $n=10$ and $n=20$ toroidal mode numbers in \figref{fig:coupmultisurf}. For example, the weak dependence of upper branch of the continuous spectrum near the BAAE gap in the vicinity of mode rational surface is a reflection of the weaker plasma response to excitations by either fluid or EP effects\cite{chen17}. By direct comparison between plots of the continuous spectrum and local dispersion curves, it is possible to identify the most important physical processes. As an example, comparing \figref{fig:coupsinglesurf} and \figref{fig:coupmultisurfn2}, we could deduce that the first three gaps observed in \figref{fig:coupmultisurfn2} with $r/a \sim 0.58$ are produced by the interaction of two ISWs. In fact, for $n=2$ and the given safety factor profile, we obtain $\nu \sim 0.5$ from Eq. (\ref{eq:356}). Furthermore, the wave poloidal mode numbers producing the gaps in \figref{fig:coupsinglesurf}  could be obtained by means of Eqs. (\ref{eq:bragg3}) and (\ref{eq:bragg4}). All this information could be routinely included in continuum spectrum plots. Nonetheless, for the sake of clarity, in this work we prefer to show only the frequency plot. Note, however, that the presence of continuous spectrum structures in a certain region of the $(r/a,\Omega)$ plane does not automatically imply continuum damping by resonant excitation\cite{chen74b,chen74a,hasegawa74} of those same structures. As anticipated in Section \ref{sec:org90d93a1}, a crucial role is played by fluctuation polarization, Eq. (\ref{eq:eigencont}), which must be computed self-consistently accounting for fluctuation structure and dispersion relation, often affected non-perturbatively by EPs\cite{chen16,zhang16,liu17,Bierwage2017,zonca14a,zonca14b,Lauber2013,chen17}. Without solving for self-consistent mode structure and dispersion relation, a qualitative information about the effective role of resonant excitation of continuum structures can still be obtained from $|e_1|^2$ adopted as {\em Alfv\'enicity parameter}
\begin{equation}
  \label{eq:alfvenicityn}
  {\cal A} = \left| e_1 (\nu, r) \right |^2 \; .
\end{equation}
With this definition, consistent with that adopted in Ref. \citenum{chen17} for quantifying mode polarization in the considered MHD limit, ${\cal A} \simeq 1$ for SAW continuum, since $g_2 \sim {\cal O} (\beta^{1/2}) g_1$ at low frequency. Meanwhile, noting  $g_1 \sim {\cal O} (\beta^{1/2}) g_2$ for ISW, ${\cal A} \sim {\cal O}(\beta)$ for the acoustic continuum. In general, $\mathcal{A}$ can be calculated for each point of the local dispersion curve and, therefore, for each point of the continuous spectrum plot. In fact, this corresponds to evaluate not only the Floquet characteristic exponent at each radial location, but the corresponding parallel fluctuation structure, $g_1^{(i)}(\eta ; \nu_i,r)$ (playing the role of ``eigenvector''), and the corresponding polarization
\begin{align}
\label{eq:corrpolariz}  
e_1^2(\nu,r) = \frac{\int (g_1^{(i)}(\eta ; \nu_i,r))^2 d \eta}{\int [ (g_1^{(i)}(\eta; \nu_i, r))^2+(g_2^{(i)} (\eta; \nu_i, r))^2] d \eta}\,,
\end{align}
consistent with Eqs. (\ref{eq:eigencont})-(\ref{eq:normpol}). This allows, in principle, to represent the Alfvénicity by means of a color-map in these graphs. A detailed analysis along these lines is beyond the scope of this work, which mainly aims at the illustration of the methodology and the generality of its applications. As a particular case, we have computed the polarization and Alfvénicity at $r/a=0.58$ for the local dispersion curves represented in \figref{fig:coupsinglesurf}. For reference, we have assumed $n=2$ such that $\nu = 0.477$ corresponds to the value on $nq(0.58)$ reduced to the first Brillouin zone (cf. Sec. \ref{sec:sawdec} ). \figref{fig:alfvenicity} illustrates the relevant continuum frequencies, for which we have computed polarization and the values of Alfvénicity, ${\cal A}$, according to Eqs. (\ref{eq:alfvenicityn})-(\ref{eq:corrpolariz}). These are, ${\cal A}=0.02, 0.06, 0.59, 0.28, 0.61, 0.34$ and $0.98$, from low to high frequencies, respectively. The first two values are very small, in consistency with the acoustic nature of the fluctuation that is predominant as expected from the green marking of the circle denoting SSW-SSW interaction in \figref{fig:coupsinglesurf}. The mixed polarization nature (orange circle, denoting SSW-SAW coupling) is also clearly indicated by the Alfvénicity, which grows for the third to the sixth branch. We could actually state that ${\cal A}$ is a proper quantitative definition of the mixed polarization in these cases. Finally, the highest frequency branch is clearly Alfvénic, as suggested by both ${\cal A} = 0.98$ and the red circle in \figref{fig:coupsinglesurf} denoting SAW-SAW coupling. In general, a SAW polarized fluctuation, characterized by $(e_1,e_2) = (1,0)$ in the uncoupled case, has weak interaction with the acoustic polarized continuum, which has $(e_1, e_2) = (0,1)$ in the uncoupled limit.
Meanwhile, an ISW polarization weakly interacts with the SAW continuous spectrum.  This point is of particular importance when inspecting the structures of continuous spectra approaching the plasma edge, where, due to the decreasing ratio of sound to Alfvén speed, high order ISW sidebands may be generated by strong shaping equilibrium modulation effects and interact with SAW. This interaction generates a thick web of continuous spectrum structures, shown in \figref{fig:coupcomparison} again for $n=2$, where they are compared with numerical results obtained by MARS, showing excellent agreement also in this case. Clearly, not all structures of this web are physically relevant, because of the important role of wave polarization. More detailed illustration of this point requires a separate analysis computing the actual Alfvén wave fluctuation spectrum in DTT for the reference scenario (see, e.g., Refs. \citenum{Wang20181,Wang2019}). 

\section{Conclusions and future perspectives}
\label{sec:org6275bbc}
This work, as an application of the general theoretical framework discussed in Refs. \citenum{chen16,zonca14a,zonca14b}, illustrates the calculation of continuous spectrum structures in realistic tokamak geometry for the MHD limiting case of governing equations\cite{chen17}. The use of ballooning formalism allows the calculation of local dispersion curves, isolating the physics information concerning the plasma response to radial local singular fluctuations, constituting the continuous spectrum. Meanwhile, toroidal and poloidal mode numbers are treated explicitly for arbitrary values, simplifying the  numerical calculation of the complicated continuous spectrum structures at high mode number and/or near the plasma boundary. The present approach clearly illustrates the importance of fluctuation polarization for the assessment of damping by resonant absorption of radial singular continuous spectrum structures. As a qualitative estimate of the coupling strength of Alfvénic fluctuations to the acoustic continuum, without solving for the actual fluctuation structures, an Alfvénicity parameter is introduced, which is related to the fluctuation polarization and can be usefully adopted to visualize the importance of acoustic couplings.

Besides the simplicity of the present approach, based on the solution of two coupled second order ordinary differential equations for the local (singular) plasma response, the interesting aspect is that the present methodology could be readily extended to 3D plasma equilibria, such as stellarators, thanks to the generality of ballooning representation\cite{dewar83}. Following the the general theoretical framework of Refs. \citenum{chen16,zonca14a,zonca14b}, the present formulation could be employed within a linear gyrokinetic description of coupled SAW and ISW continua. Relevant equations, in this case, would be the short radial scale gyrokinetic vorticity equation (Eq. (6) of Ref. \citenum{zonca14b}) and the quasineutrality condition (Eq. (36) of the same work). Although the uderlying equations are different, the fundamental structure would be that of two second order periodic ordinary differential equations for the local (singular) plasma response, to be analyzed according to the well-known Floquet theory.

Within the present framework, the calculation of continuous spectrum, corresponding parallel fluctuation structures and polarization vectors provides the boundary conditions for computing any physical mode structures and corresponding dispersion relations in arbitrary geometry.
\section*{Acknowledgment}
\label{sec:ack}
This work has been carried out within the framework of the EUROfusion Consortium and has received funding from the Euratom research and training programme 2014-2018 and 2019-2020 under grant agreement No 633053. The views and opinions expressed herein do not necessarily reflect those of the European Commission.

We thank DTT Team for the reference scenario.

\appendix

\section{Derivation of useful integral quadratic forms}
\label{app:A}
When we take the decoupled ($\kappa_g=0$) limit of the system of differential equations, Eqs. (\ref{eq:36}) and (\ref{condhgvat}), it is possible to 
derive the following integral quadratic forms
\begin{eqnarray}
& & \int d \eta \left| \partial_\eta g_1 \right|^2  = \hat \rho_{m0} \Omega^2 \int d \eta \hat J_\eta^2 \left| g_1 \right|^2  \; , \nonumber \\
& & \int d \eta \left| \partial_\eta \left( | \bm \nabla r | g_2 \right) \right|^2 \frac{1}{| \bm \nabla r |^2} = \frac{ 2 \hat \rho_{m0} \Omega^2}{\Gamma \bar \beta} \int d \eta \hat J_\eta^2 \hat B_0^2 \left| g_2 \right|^2  \; . \label{eq:quadform}
\end{eqnarray}
In the neighborhood of a given point of intersecting SAW and ISW continua, identified by a given $\nu(\Omega)$ for fixed $r/a$, we can
write $$ \int d \eta \left| \partial_\eta g_1 \right|^2  = (nq - m)^2  \int d \eta \left| g_1 \right|^2  \; ;$$ and, similarly,
$$ \int d \eta \left| \partial_\eta \left( | \bm \nabla r | g_2 \right) \right|^2 \frac{1}{| \bm \nabla r |^2} = (nq - m - p)^2  \int d \eta \left| g_2 \right|^2 \; ,$$ with $p \in \mathbb Z$.
Here, we have noted that, near the given point of intersecting SAW and ISW continua, the $g_1$ and $g_2$ solutions are modulated
by the equilibrium non-uniformity within the flux surface. The first of Eqs. (\ref{eq:quadform}) is what allows deriving Eq. (\ref{eq:SAWgap}), while the combination
of the two is what readily yields Eqs. (\ref{eq:bragg3}) and (\ref{eq:bragg4}), with the following definition of $\hat \beta$:
\begin{equation}
\hat \beta \equiv \bar \beta \frac{ \int d \eta \hat J_\eta^2 \left| g_1 \right|^2 }{ \int d \eta \left| g_1 \right|^2 } \frac{ \int  d \eta \left| g_2 \right|^2}{ \int d \eta \hat J_\eta^2 \hat B_0^2 \left| g_2 \right|^2 } \; . \label{eq:hatbeta}
\end{equation}

\end{document}